# Generalized Distributive Law for ML Decoding of Space-Time Block Codes

Lakshmi Prasad Natarajan and B. Sundar Rajan, *Senior Member, IEEE*

*Abstract*—The problem of designing good Space-Time Block Codes (STBCs) with low maximum-likelihood (ML) decoding complexity has gathered much attention in the literature. All the known low ML decoding complexity techniques utilize the same approach of exploiting either the multigroup decodable or the fast-decodable (conditionally multigroup decodable) structure of a code. We refer to this well known technique of decoding STBCs as *Conditional ML (CML) decoding*. In this paper we introduce a new framework to construct ML decoders for STBCs based on the Generalized Distributive Law (GDL) and the Factor-graph based Sum-Product Algorithm. We say that an STBC is *fast GDL decodable* if the order of GDL decoding complexity of the code is strictly less than $M^\lambda$, where $\lambda$ is the number of independent symbols in the STBC, and $M$ is the constellation size. We give sufficient conditions for an STBC to admit fast GDL decoding, and show that both multigroup and conditionally multigroup decodable codes are fast GDL decodable. For any STBC, whether fast GDL decodable or not, we show that the GDL decoding complexity is strictly less than the CML decoding complexity. For instance, for any STBC obtained from Cyclic Division Algebras which is not multigroup or conditionally multigroup decodable, the GDL decoder provides about 12 times reduction in complexity compared to the CML decoder. Similarly, for the Golden code, which is conditionally multigroup decodable, the GDL decoder is only half as complex as the CML decoder.

## I. INTRODUCTION

THE complexity with which a Space-Time Block Code (STBC) can be maximum-likelihood (ML) decoded is an important parameter from an implementation point of view. Consequently, the problem of designing codes with high rate and good error performance that admit low complexity ML decoding is of much interest in the literature. This problem was first attacked by constructing *multigroup decodable codes* which have the property that the information symbols of the code can be partitioned into several groups, and each group of symbols can be ML decoded independent of other symbol groups. Examples include the Orthogonal Designs [1]–[3] and the higher rate multigroup decodable STBCs constructed in [4]–[15]. In [16], it was shown that a new class of STBCs called *fast-decodable* or *conditionally multigroup decodable codes* allow reduced complexity decoding as well. These codes contain a lower rate multigroup decodable STBC as a subcode, and this property is leveraged to decode such STBCs with low complexity. Examples of fast-decodable codes available in the literature include [17]–[24], the Silver code [25], [26] and the Golden Code [27]–[29], [18]. All known low complexity ML decoders have the same unified approach of exploiting either the multigroup decodability or the conditional multigroup decodability of a code. This method is well known and widely used in the literature, and we will refer to it as *Conditional ML (CML)* decoding.

The Generalized Distributive Law [30] and its equivalent, factor graph based approach, known as the Sum-Product Algorithm [31] are message-passing algorithms that efficiently solve a class of computation problems called Marginalize a Product Function (MPF) problems. The Generalized Distributive Law (GDL) includes as special cases the Viterbi's algorithm [32], the BCJR algorithm [33], the Fast-Fourier Transform [34], the Turbo [35] and LDPC decoding algorithms [36], [37]. In this paper, we first identify that the ML decoding problem of any STBC is equivalent to the problem of minimizing a multivariate, second degree real polynomial, where the variables assume values from a finite signal set. Using this observation we show that the ML decoding of any STBC is an MPF problem, and hence, the GDL is a natural choice for constructing low complexity ML decoders. The contribution and organization of this paper are as follows.

- We introduce a new, GDL based framework to design ML decoders for STBCs (Section III and Section IV-A). Since the GDL is computationally efficient, this new framework provides a rich scope for designing low complexity ML decoders.
- We show that the GDL decoding complexity of any code is strictly less than its CML decoding complexity (Theorems 2 and 3, Section V-B). As an application of our results, we show that for any STBC obtained from Cyclic Division Algebras [38] which is not multigroup or conditionally multigroup decodable, the GDL decoder is approximately 12 times less complex than the CML decoder. In case of the Golden code, which is conditionally multigroup decodable, the GDL decoder is roughly half as complex as the CML decoder (Example G.4, Section V-C). The GDL can lead to reductions in the order of decoding complexity as well, when compared to the CML decoder. We give explicit examples of two classes of STBCs, the Toeplitz codes [39] and the Overlapped Alamouti Codes [40], where the GDL decoder has a lower complexity order than the CML decoder (Section V-B).
- We give sufficient conditions for a code to be *fast GDL decodable* i.e., to admit low complexity GDL decod-

This work was supported partly by the DRDO-IISc program on Advanced Research in Mathematical Engineering through a research grant, and partly by the INAE Chair Professorship grant to B. S. Rajan. The material in this paper was presented in part at the IEEE Information Theory Workshop, Paraty, Brazil, October 16-20, 2011.

The authors are with the Department of Electrical Communication Engineering, Indian Institute of Science, Bangalore-560012, India (e-mail: {nlp, bsrajan}@ece.iisc.ernet.in).

ing, and show that both multigroup and conditionally multigroup decodable codes are amenable to fast GDL decoding (Section IV-B). Using the new GDL framework we also provide tools to readily identify multigroup and conditionally multigroup decodable codes (Section IV-B).

- When the information symbols of a code are encoded using a PAM signal set, we show that the GDL algorithm can exploit the structure of PAM to lead to further reduction in decoding complexity (Section V-C).

A brief review of the GDL is given in Section II, and the paper is concluded in Section VI.

*Notations* - Throughout the paper, matrices (vectors) are denoted by bold, uppercase (lowercase) letters. The Hermitian and Frobenius norm of a matrix $\mathbf{X}$ are denoted by $\mathbf{X}^H$ and $||\mathbf{X}||$ respectively. For a square matrix $\mathbf{X}$, $tr(\mathbf{X})$ denotes the trace of $\mathbf{X}$. Unless used as a subscript or to denote indices, $j$ represents $\sqrt{-1}$. The set of all real and complex numbers are denoted by $\mathbb{R}$ and $\mathbb{C}$, respectively. The $m \times m$ sized null matrix is denoted by $\mathbf{O}_m$. For any set $\mathcal{I}$, its complement in the corresponding universal set is denoted by $\mathcal{I}^c$.

## II. A Brief Review of the Generalized Distributive Law

In Section III we show that the ML decoding of STBCs is an instance of a particular class of MPF problems: the MPF problems on the min-sum semiring over the real numbers $\mathbb{R}$. We now recall the definition of this class of computational problems, their GDL solution and some properties of the GDL which we use in the later sections.

### A. MPF problems on the min-sum semiring over $\mathbb{R}$

Consider the union of the set of real numbers $\mathbb{R}$ and the element infinity, $\infty$. With *multiplication* defined on this set as the sum of two elements, and *addition* defined as the operation of taking the minimum, we get the min-sum semiring over $\mathbb{R}$. The elements $\infty$ and $0$ are the *additive* and *multiplicative identities* respectively. The class of MPF problems defined on this semiring are as follows [30]. Let $\mathbf{x}_1, \ldots, \mathbf{x}_N$ be variables that take values independently from finite sets $\mathcal{A}_1, \ldots, \mathcal{A}_N$ respectively. For any $\mathcal{I} = \{i_1, \ldots, i_{|\mathcal{I}|}\} \subset \{1, \ldots, N\}$ with $i_1 < i_2 < \cdots < i_{|\mathcal{I}|}$, denote by $\mathcal{A}_\mathcal{I}$ the set $\mathcal{A}_{i_1} \times \cdots \times \mathcal{A}_{i_{|\mathcal{I}|}}$, and denote by $\mathbf{x}_\mathcal{I}$ the variable list $(\mathbf{x}_{i_1}, \ldots, \mathbf{x}_{i_{|\mathcal{I}|}})$. Let $\mathcal{S} = \{\mathcal{I}_1, dots, \mathcal{I}_L\}$ be a set of $L$ subsets of $\{1, \ldots, N\}$, and for each $\ell = 1, \ldots, L$, let $\alpha_\ell : \mathcal{A}_{\mathcal{I}_\ell} \to \mathbb{R}$ be a given function i.e., a table of values. Define functions $\beta : \mathcal{A}_{\{1,\ldots,N\}} \to \mathbb{R}$ and $\beta_\ell : \mathcal{A}_{\mathcal{I}_\ell} \to \mathbb{R}$, $\ell = 1, \ldots, L$, as follows:

$$\beta(\mathbf{x}_1, \ldots, \mathbf{x}_N) = \sum_{\ell=1}^L \alpha_\ell(\mathbf{x}_{\mathcal{I}_\ell}) \text{ and} \quad (1)$$

$$\beta_\ell(\mathbf{x}_{\mathcal{I}_\ell}) = \min_{\mathbf{x}_{\mathcal{I}_\ell^c} \in \mathcal{A}_{\mathcal{I}_\ell^c}} \beta(\mathbf{x}_1, \ldots, \mathbf{x}_N), \quad (2)$$

where $\sum$ denotes addition of real numbers, and $\mathcal{I}_\ell^c$ is the complement of $\mathcal{I}_\ell$ in $\{1, \ldots, N\}$. The MPF problem on the min-sum semiring over $\mathbb{R}$ is to compute the table of values of the function $\beta_\ell$ for one or more $\ell = 1, \ldots, L$, given the functions $\alpha_1, \ldots, \alpha_L$. The function $\beta$ is called the *global kernel* and the function $\beta_\ell$ is called the $\mathbf{x}_{\mathcal{I}_\ell}$-*marginalization of* $\beta$ or the *objective function at* $\mathcal{I}_\ell$.

### B. The Generalized Distributive Law

The GDL is a message-passing algorithm that operates on a simple tree (an undirected, unweighted, connected[1] graph with no loops, cycles or multiple edges) $\mathcal{G} = (\mathcal{V}, \mathcal{E})$. Each vertex $v \in \mathcal{V}$ is associated with a function $\alpha_v : \mathcal{A}_{\mathcal{I}_v} \to \mathbb{R}$, for some $\mathcal{I}_v \subset \{1, \ldots, N\}$. The function $\alpha_v$ is called the *local kernel* at $v$, and the variable list $\mathbf{x}_{\mathcal{I}_v}$ is called the *local domain* at $v$. The tree $\mathcal{G}$ can be used to solve the MPF problem given in (2) using the GDL if it satisfies the following three conditions:

C.1 for each $\ell = 1, \ldots, L$, there exists a $v \in \mathcal{V}$ such that $\mathcal{I}_\ell = \mathcal{I}_v$,
C.2 the global kernel $\beta = \sum_{\ell=1}^L \alpha_\ell = \sum_{v \in \mathcal{V}} \alpha_v$, and
C.3 the tree $\mathcal{G}$ satisfies the *junction tree condition*, i.e., for each $n = 1, \ldots, N$, the subgraph of $\mathcal{G}$ consisting of those vertices whose local domains contain $\mathbf{x}_n$ together with the edges connecting these vertices is connected.

A tree $\mathcal{G}$ that satisfies all the three conditions above is said to be a *junction tree* for the given MPF problem. In general there is no unique junction tree for an MPF problem, and different junction trees may lead to GDL algorithms with varying complexities of implementation. Various methods to construct/transform junction trees are given in [30], [31].

For any two neighboring vertices $u$ and $v$, the *directed message* from $u$ to $v$ is a table of values of a function $\mu_{u,v} : \mathcal{A}_{\mathcal{I}_u \cap \mathcal{I}_v} \to \mathbb{R}$. To send a message to $v$, the vertex $u$ forms the sum of its local kernel with the messages that it has received from all its neighbors other than $v$, and then marginalizes this sum with respect to the variables common to $u$ and $v$, i.e.,

$$\mu_{u,v}(\mathbf{x}_{\mathcal{I}_u \cap \mathcal{I}_v}) = \min_{\mathbf{x}_{\mathcal{I}_u \setminus \mathcal{I}_v}} \left( \alpha_u(\mathbf{x}_{\mathcal{I}_u}) + \sum_{\substack{w \ adj \ u \\ w \neq v}} \mu_{w,u}(\mathbf{x}_{\mathcal{I}_w \cap \mathcal{I}_u}) \right),$$

where $w \ adj \ u$ denotes that the vertices $w$ and $u$ are neighbors. The *state* of the vertex $u$ is a table of values of a function $\sigma_u : \mathcal{A}_{\mathcal{I}_u} \to \mathbb{R}$. Initially $\sigma_u$ is set to be equal to the local kernel at $u$. During the GDL algorithm it is updated as the sum of the local kernel at $u$ with the messages that $u$ has received from all its neighbors, i.e.,

$$\sigma_u(\mathbf{x}_{\mathcal{I}_u}) = \alpha_u(\mathbf{x}_{\mathcal{I}_u}) + \sum_{w \ adj \ u} \mu_{w,u}(\mathbf{x}_{\mathcal{I}_w \cap \mathcal{I}_u}).$$

In order to solve the *all-vertex problem*, i.e., to compute the $\mathbf{x}_{\mathcal{I}_v}$-marginalization of $\beta$ for every $v \in \mathcal{V}$, every vertex is made to send a message to a neighbor when for the first time it receives messages from all its other neighbors. So the messages begin at the leaves of the junction tree, proceed inwards into the tree and then travel back outwards. At the end of this message-passing schedule, each vertex computes its state, which is guaranteed to be equal to the objective function at that vertex [30]. The objective function $\beta_\ell$ given in (2) is thus equal to the state of any vertex $v$ with $\mathcal{I}_v = \mathcal{I}_\ell$. To solve a *single-vertex problem*, i.e., to compute the $\mathbf{x}_{\mathcal{I}_v}$-marginalization of $\beta$ for a given vertex $v$, all the edges of the junction tree are directed towards the *root* $v$. Every vertex except $v$ sends exactly one message to its neighbor along the unique path to $v$ when it has received messages from all its other neighbors. The state at $v$ is computed once $v$ receives messages from all its neighbors, and this equals the objective function at $v$.

---
[1]A graph is said to be connected if there exists a path between every pair of nodes.

The total number of additions and pairwise comparisons (for implementing $\min$) in the case of single-vertex problem for any root vertex $v$ is equal to

$$\begin{aligned} \mathsf{C}(\mathcal{G}) &= \sum_{u \in \mathcal{V}} d_u |\mathcal{A}_{\mathcal{I}_u}| - \sum_{(w,u) \in \mathcal{E}} |\mathcal{A}_{\mathcal{I}_w \cap \mathcal{I}_u}| \\ &= \sum_{(w,u) \in \mathcal{E}} \left( |\mathcal{A}_{\mathcal{I}_w}| + |\mathcal{A}_{\mathcal{I}_u}| - |\mathcal{A}_{\mathcal{I}_w \cap \mathcal{I}_u}| \right), \end{aligned} \quad (3)$$

where $d_u$ is the degree of the vertex $u$. The all-vertex GDL schedule can be implemented with complexity of at the most $4\mathsf{C}(\mathcal{G})$. The complexity order for both single and all-vertex problems is thus $\max_{u \in \mathcal{V}} |\mathcal{A}_{\mathcal{I}_u}|$.

The messages passed during the GDL schedule can be characterized precisely using the local kernels of $\mathcal{G}$. In both the single and the all-vertex GDL schedules, the directed message from a vertex $u$ to its neighbor $v$ is the $\mathbf{x}_{\mathcal{I}_u \cap \mathcal{I}_v}$-marginalization of the sum of the local kernels of all the vertices descending from $u$ [31]. More formally, consider the two disjoint trees $\mathcal{G}_{u \setminus v}$ and $\mathcal{G}_{v \setminus u}$ obtained from $\mathcal{G}$ by removing the edge $(u,v) \in \mathcal{E}$, such that $\mathcal{G}_{u \setminus v}$ contains the vertex $u$ and $\mathcal{G}_{v \setminus u}$ contains $v$. Then we have

$$\mu_{u,v}(\mathbf{x}_{\mathcal{I}_u \cap \mathcal{I}_v}) = \min_{\mathbf{x}_{(\mathcal{I}_u \cap \mathcal{I}_v)^c}} \sum_{w \in \mathcal{G}_{u \setminus v}} \alpha_w(\mathbf{x}_{\mathcal{I}_w}).$$

The GDL algorithm capitalizes on the 'factorization' of $\beta$, as given in (1), into $L$ functions whose domains are smaller than that of $\beta$ itself, and hence are less complex to work with compared to $\beta$. During the message-passing, partial sums of these 'smaller' functions are calculated, and these are used efficiently to compute the various $\mathbf{x}_{\mathcal{I}_\ell}$-marginalizations of $\beta$.

## III. THE GDL DECODING OF SPACE-TIME BLOCK CODES

In this section, we first introduce the notion of *encoding groups* in STBCs obtained from linear designs, and then using this concept, formulate the ML decoding of such STBCs as an MPF problem over the min-sum semiring over $\mathbb{R}$. We then propose a junction tree to decode any STBC obtained from linear designs using the GDL message-passing algorithm.

### A. Channel model, designs and encoding groups

We consider the block fading MIMO channel with full channel state information (CSI) at the receiver and no CSI at the transmitter. For an $n_t \times n_r$ MIMO transmission, we have

$$\mathbf{Y} = \mathbf{H}\mathbf{X} + \mathbf{N}, \quad (4)$$

where $\mathbf{X} \in \mathbb{C}^{n_t \times T}$ is the codeword matrix transmitted over $T$ channel uses, $\mathbf{N} \in \mathbb{C}^{n_r \times T}$ is a complex white Gaussian noise matrix whose entries are i.i.d. with zero mean and unit variance, and $\mathbf{H} \in \mathbb{C}^{n_r \times n_t}$ is the channel matrix with arbitrary probability distribution. An STBC $\mathcal{C}$ is a finite set of $n_t \times T$ complex matrices. We consider codes that are obtained from designs $\mathbf{S} = \sum_{i=1}^{K} s_i \mathbf{A}_i$, where $s_1, \ldots, s_K$ are real variables or *information symbols* and $\mathbf{A}_i \in \mathbb{C}^{n_t \times T}$ are the *weight* or *linear dispersion* matrices [12], [41]. The rate of the resulting code is $\frac{K}{2T}$ complex symbols per channel use. Commonly in the literature the real variables $\{s_i\}$ are combined pairwise, and the design is represented in terms of the resulting complex information symbols. Examples include matrix designs whose individual entries are complex linear combinations of complex variables and their conjugates.

Let the symbols $\{s_1, \ldots, s_K\}$ be partitioned into $N$ subsets, called *encoding groups*, such that the symbols in different encoding groups are encoded independently and all the symbols in each encoding group are encoded jointly. For $n = 1, \ldots, N$, let $\mathbf{x}_n$ be the vector consisting of the information symbols belonging to the $n^{th}$ encoding group, and let $\mathbf{x}_n$ be encoded using a finite set $\mathcal{A}_n \subset \mathbb{R}^{\lambda_n}$, where $\lambda_n$ is the number of real symbols in the $n^{th}$ encoding group. The STBC obtained from the design $\mathbf{S}$ and the signal sets $\mathcal{A}_1, \ldots, \mathcal{A}_N$ is

$$\mathcal{C} = \left\{ \sum_{i=1}^{K} s_i \mathbf{A}_i \;\middle|\; \mathbf{x}_n \in \mathcal{A}_n, \; n = 1, \ldots, N \right\}.$$

*Example T.1:* Consider the Toeplitz code [39] for $n_t = 2$ antennas and $T = 10$ time slots. The number of real symbols $K = 18$ and the design $\mathbf{S} =$

$$\begin{bmatrix} s_1 + js_2 & s_3 + js_4 & s_5 + js_6 & \cdots & s_{17} + js_{18} & 0 \\ 0 & s_1 + js_2 & s_3 + js_4 & \cdots & s_{15} + js_{16} & s_{17} + js_{18} \end{bmatrix}.$$

Let the complex symbols $s_{2n-1} + js_{2n}$, $n = 1, \ldots, 9$, be encoded using a HEX constellation [42] $\mathcal{A}_{HEX} \subset \mathbb{R}^2$. This STBC has $N = 9$ encoding groups and the vectors $\mathbf{x}_n$, $n = 1, \ldots, 9$, are given by $\mathbf{x}_n = \begin{bmatrix} s_{2n-1} & s_{2n} \end{bmatrix}^T$. The number of symbols per each encoding group is $\lambda_n = 2$ and the finite sets $\mathcal{A}_n = \mathcal{A}_{HEX}$ for $n = 1, \ldots, 9$. ∎

A subset of real information symbols $\{s_1, \ldots, s_K\}$ that are encoded together using an arbitrary joint signal set must be decoded jointly by an ML decoder. The encoding groups $\mathbf{x}_1, \ldots, \mathbf{x}_N$ are the fundamental units of information variables that any ML decoder will operate on. For a given STBC the choice of the weight matrices $\{\mathbf{A}_i\}$, encoding groups $\{\mathbf{x}_n\}$ and the signal sets $\{\mathcal{A}_n\}$ may not be unique. As illustrated in the following example, a careful choice of the weight matrices and signal sets can reduce the number of real symbols per encoding group. This reduction in encoding complexity may get reflected as a reduction in the ML decoding complexity at the receiver.

*Example G.1:* Consider the Dayal-Varanasi version of the Golden Code [28]:

$$\mathbf{S}_1 = \begin{bmatrix} s_1 + js_2 & \gamma(s_5 + js_6) \\ \gamma(s_7 + js_8) & s_3 + js_4 \end{bmatrix},$$

where $\gamma = \sqrt{-j}$ and the symbol vectors $\begin{bmatrix} s_1 + js_2 & s_3 + js_4 \end{bmatrix}^T$ and $\begin{bmatrix} s_5 + js_6 & s_7 + js_8 \end{bmatrix}^T$ are encoded independently using a constellation from the rotated lattice $R\mathbb{Z}[j]^2$ with

$$R = \begin{bmatrix} c & s \\ -s & c \end{bmatrix}, \quad c = cos\left(\frac{tan^{-1}(2)}{2}\right) \text{ and}$$
$$s = sin\left(\frac{tan^{-1}(2)}{2}\right).$$

A naive choice for the symbol groups is

$$\mathbf{x}_1 = \begin{bmatrix} s_1 & s_2 & s_3 & s_4 \end{bmatrix}^T, \mathbf{x}_2 = \begin{bmatrix} s_5 & s_6 & s_7 & s_8 \end{bmatrix}^T.$$

The corresponding weight matrices are

$$\mathbf{A}_1 = \begin{bmatrix} 1 & 0 \\ 0 & 0 \end{bmatrix}, \mathbf{A}_2 = \begin{bmatrix} j & 0 \\ 0 & 0 \end{bmatrix}, \mathbf{A}_3 = \begin{bmatrix} 0 & 0 \\ 0 & 1 \end{bmatrix},$$
$$\mathbf{A}_4 = \begin{bmatrix} 0 & 0 \\ 0 & j \end{bmatrix}, \mathbf{A}_5 = \begin{bmatrix} 0 & \gamma \\ 0 & 0 \end{bmatrix}, \mathbf{A}_6 = \begin{bmatrix} 0 & j\gamma \\ 0 & 0 \end{bmatrix},$$
$$\mathbf{A}_7 = \begin{bmatrix} 0 & 0 \\ \gamma & 0 \end{bmatrix} \text{ and } \mathbf{A}_8 = \begin{bmatrix} 0 & 0 \\ j\gamma & 0 \end{bmatrix}.$$

It is shown in Example G.4 of Section V-C that this choice of encoding groups leads to GDL based decoders with complexity equal to that of brute-force ML decoding. A better choice of weight matrices and encoding groups can be obtained by a simple linear transformation of the symbols $\{s_i\}$. The resulting design $\mathbf{S_2}$ is given in (5) at the top of the next page. The symbols $\{s_i\}$ of this new design are encoded independently of each other using a PAM constellation. Both $\mathbf{S_1}$ and $\mathbf{S_2}$ give the same STBC though they are encoded using different sets of weight matrices and constellations. The number of encoding groups in $\mathbf{S_2}$ is 8, and each symbol $s_i$ forms an encoding group by itself, i.e., $\mathbf{x}_n = [s_n]$, $n = 1, \ldots, 8$. The corresponding weight matrices are

$$\mathbf{A}_1 = \begin{bmatrix} c & 0 \\ 0 & -s \end{bmatrix}, \mathbf{A}_2 = \begin{bmatrix} jc & 0 \\ 0 & -js \end{bmatrix}, \mathbf{A}_3 = \begin{bmatrix} s & 0 \\ 0 & c \end{bmatrix},$$
$$\mathbf{A}_4 = \begin{bmatrix} js & 0 \\ 0 & jc \end{bmatrix}, \mathbf{A}_5 = \begin{bmatrix} 0 & \gamma c \\ -\gamma s & 0 \end{bmatrix}, \mathbf{A}_6 = \begin{bmatrix} 0 & j\gamma c \\ -j\gamma s & 0 \end{bmatrix},$$
$$\mathbf{A}_7 = \begin{bmatrix} 0 & \gamma s \\ \gamma c & 0 \end{bmatrix} \text{ and } \mathbf{A}_8 = \begin{bmatrix} 0 & j\gamma s \\ j\gamma c & 0 \end{bmatrix}.$$

This choice of encoding groups leads to reduced complexity ML decoding as will be shown in Example G.4. ■

### B. The GDL Decoding of STBCs

Given the $n_r \times T$ received matrix $\mathbf{Y}$ in (4), the ML decoder finds the set of variables $\{s_1, \ldots, s_K\}$ that minimizes $||\mathbf{Y} - \mathbf{H}\sum_{i=1}^K s_i \mathbf{A}_i||^2$. The ML decoding problem is to find

$$\arg\min tr\left((\mathbf{Y} - \sum_{i=1}^K s_i \mathbf{H A}_i)(\mathbf{Y}^H - \sum_{i=1}^K s_i \mathbf{A}_i^H \mathbf{H}^H)\right)$$
$$= \arg\min tr(\mathbf{YY}^H) + \sum_{i=1}^K s_i tr(-\mathbf{HA}_i \mathbf{Y}^H - \mathbf{YA}_i^H \mathbf{H}^H)$$
$$+ \sum_{i=1}^K s_i^2 tr(\mathbf{HA}_i \mathbf{A}_i^H \mathbf{H}^H)$$
$$+ \sum_{i=1}^K \sum_{j>i} s_i s_j tr(\mathbf{H}(\mathbf{A}_i \mathbf{A}_j^H + \mathbf{A}_j \mathbf{A}_i^H)\mathbf{H}^H)$$
$$= \arg\min f(s_1, \ldots, s_K),$$

where $tr(\cdot)$ is the trace of a square matrix, and

$$f(s_1, \ldots, s_K) = \sum_{i=1}^K (s_i \xi_i + s_i^2 \xi_{i,i}) + \sum_{j>i} s_i s_j \xi_{i,j},$$
$$\xi_i = tr(-\mathbf{HA}_i \mathbf{Y}^H - \mathbf{YA}_i^H \mathbf{H}^H),$$
$$\xi_{i,j} = tr(\mathbf{H}(\mathbf{A}_i \mathbf{A}_j^H + \mathbf{A}_j \mathbf{A}_i^H)\mathbf{H}^H) \text{ for } j > i, \text{ and}$$
$$\xi_{i,i} = tr(\mathbf{HA}_i \mathbf{A}_i^H \mathbf{H}^H).$$

Since the matrices $\mathbf{HA}_i \mathbf{Y}^H + \mathbf{YA}_i^H \mathbf{H}^H$, $\mathbf{HA}_i \mathbf{A}_i^H \mathbf{H}^H$ and $\mathbf{H}(\mathbf{A}_i \mathbf{A}_j^H + \mathbf{A}_j \mathbf{A}_i^H)\mathbf{H}^H$ are Hermitian, the coefficients $\xi_i$, $\xi_{i,i}$, $\xi_{i,j}$ are all real.

The function $f(s_1, \ldots, s_K)$ is a second degree polynomial over $\mathbb{R}$. We now partition the terms of this polynomial according to the encoding groups $\{\mathbf{x}_n\}$. The terms in $f$ that consist of variables only from the $n^{th}$ encoding group are summed together into the function $\alpha_n(\mathbf{x}_n)$. For $n < m$, those terms in $f$ that contain exactly one variable each from the $n^{th}$ and the $m^{th}$ encoding groups are summed together to get the function $\alpha_{n,m}(\mathbf{x}_n, \mathbf{x}_m)$. For $n = 1, \ldots, N$, let $\psi(n)$ denote the set of indices of those real symbols $s_i$ that are in the $n^{th}$ encoding group $\mathbf{x}_n$. Then for $n = 1, \ldots, N$, we have

$$\alpha_n(\mathbf{x}_n) = \sum_{i \in \psi(n)} \left(s_i \xi_i + s_i^2 \xi_{i,i}\right) + \sum_{\substack{j>i \\ i,j \in \psi(n)}} s_i s_j \xi_{i,j},$$

and for all $1 \leq n < m \leq N$ we have

$$\alpha_{n,m}(\mathbf{x}_n, \mathbf{x}_m) = \sum_{\substack{i \in \psi(n) \\ j \in \psi(m)}} s_i s_j \xi_{i,j}. \quad (6)$$

Define

$$\beta(\mathbf{x}_1, \ldots, \mathbf{x}_N) = \sum_{n=1}^N \alpha_n(\mathbf{x}_n) + \sum_{m>n} \alpha_{n,m}(\mathbf{x}_n, \mathbf{x}_m). \quad (7)$$

By definition, $\beta(\mathbf{x}_1, \ldots, \mathbf{x}_N) = f(s_1, \ldots, s_K)$ and the ML solution is $(\hat{\mathbf{x}}_1, \ldots, \hat{\mathbf{x}}_N) = \arg\min \beta(\mathbf{x}_1, \ldots, \mathbf{x}_N)$. If the ML solution is unique then for each $n = 1, \ldots, N$, we have $\hat{\mathbf{x}}_n = \arg\min \beta_n(\mathbf{x}_n)$ where

$$\beta_n(\mathbf{x}_n) = \min_{\mathbf{x}_{\{n\}^c} \in \mathcal{A}_{\{n\}^c}} \beta(\mathbf{x}_1, \ldots, \mathbf{x}_N). \quad (8)$$

The definition of $\beta$ in (7) provides a natural 'factorization' of the global kernel in terms of the functions $\alpha_n$ and $\alpha_{n,m}$ whose domains are much smaller than that of $\beta$, and hence

$$\mathbf{S_2} = \begin{bmatrix} s_1c + s_3s + js_2c + js_4s & \gamma(s_5c + s_7s + js_6c + js_8s) \\ \gamma(-s_5s + s_7c - js_6s + js_8c) & -s_1s + s_3c - js_2s + js_4c \end{bmatrix} \quad (5)$$

are easier to compute. From (2) and (8), we see that the ML decoding of an STBC is an MPF problem, and hence it can be solved using the GDL which efficiently processes the partial sums of $\alpha_n$, $\alpha_{n,m}$ to compute the $\mathbf{x}_n$-marginalizations of $\beta$. The ML solution for $\mathbf{x}_n$ can be obtained by first computing the $\mathbf{x}_n$-marginalization of the global kernel $\beta$ in (8) and then finding the argument $\mathbf{x}_n$ that minimizes $\beta_n$.

When the ML solution is not unique an arbitration is required after solving the MPF problem. To illustrate this, consider the case $N = 2$ and say both $(\hat{\mathbf{x}}_1, \hat{\mathbf{x}}_2) = (\mathbf{a}_1, \mathbf{a}_2)$ and $(\hat{\mathbf{x}}_1, \hat{\mathbf{x}}_2) = (\mathbf{b}_1, \mathbf{b}_2)$ are ML solutions. On solving the MPF problem (8) we would obtain a table of values for the functions $\beta_1(\mathbf{x}_1)$ and $\beta_2(\mathbf{x}_2)$. However, both $\mathbf{a}_1$ and $\mathbf{a}_2$ minimize $\beta_1$, and both $\mathbf{b}_1$ and $\mathbf{b}_2$ minimize $\beta_2$. Thus we only know that the ML solutions belong to the set $\{(\mathbf{a}_1, \mathbf{a}_2), (\mathbf{a}_1, \mathbf{b}_2), (\mathbf{b}_1, \mathbf{a}_2), (\mathbf{b}_1, \mathbf{b}_2)\}$. In order to obtain the ML solutions, the ML metric $||\mathbf{Y} - \mathbf{HX}||^2$ for each of these tuples should be calculated. The following lemma says that for an i.i.d. Rayleigh fading channel the ML solution of an STBC is unique with probability 1, and hence this arbitration step can be safely ignored.

*Lemma 1:* Let $\mathcal{C}$ be any STBC, and let the entries of the channel matrix $\mathbf{H}$ be i.i.d. complex Gaussian random variables with zero mean and unit variance. Then with probability 1 the ML solution for the transmitted codeword for the channel (4) is unique.

*Proof:* Let $\mathbf{X}_1$ and $\mathbf{X}_2$ be two distinct codewords. We will first show that with probability (w.p.) 1 $\mathbf{HX}_1 \neq \mathbf{HX}_2$, and then show that given $\mathbf{HX}_1 \neq \mathbf{HX}_2$ the probability that both $\mathbf{X}_1$ and $\mathbf{X}_2$ are ML solutions is 0. Since $\mathbf{X}_1 \neq \mathbf{X}_2$, there exists a column of $(\mathbf{X}_1 - \mathbf{X}_2)$ which is non-zero. Suppose the $j^{th}$ column of $(\mathbf{X}_1 - \mathbf{X}_2)$ is non-zero, the $(1, j)^{th}$ entry of the matrix $\mathbf{H}(\mathbf{X}_1 - \mathbf{X}_2)$ is a complex Gaussian random variable with zero mean and non-zero variance. Then the $(1, j)^{th}$ entry of $\mathbf{H}(\mathbf{X}_1 - \mathbf{X}_2)$ is non-zero w.p. 1 and hence $\mathbf{HX}_1 \neq \mathbf{HX}_2$ w.p. 1.

Now suppose $\mathbf{X}_0$ is the transmitted codeword and $\mathbf{H}$ is such that $\mathbf{HX}_1 \neq \mathbf{HX}_2$. Let $vec(\cdot)$ denote the vectorization of a matrix. Then $vec(\mathbf{H}(\mathbf{X}_0 - \mathbf{X}_1)) \neq vec(\mathbf{H}(\mathbf{X}_0 - \mathbf{X}_2))$. Both $\mathbf{X}_1$ and $\mathbf{X}_2$ will be ML solutions only if the $n_rT$-dimensional white Gaussian noise vector $vec(\mathbf{N})$ belongs to the the set of points in $\mathbb{C}^{n_rT}$ that are equidistant from $vec(\mathbf{H}(\mathbf{X}_0 - \mathbf{X}_1))$ and $vec(\mathbf{H}(\mathbf{X}_0 - \mathbf{X}_2))$. Since $vec(\mathbf{H}(\mathbf{X}_0 - \mathbf{X}_1)) \neq vec(\mathbf{H}(\mathbf{X}_0 - \mathbf{X}_2))$, this set is a coset of an $(n_rT - 1)$-dimensional subspace of $\mathbb{C}^{n_rT}$ and the probability that $vec(\mathbf{N})$ belongs to this hyperplane is 0. This completes the proof. ∎

A junction tree to solve the MPF problem (8) is shown in Fig. 1. The tree can be viewed as consisting of three sections. At the center of the tree is the *core* consisting of only the $(\mathbf{x}_1, \ldots, \mathbf{x}_N)$ vertex. The core is surrounded by *tier 1*: a layer of $(\mathbf{x}_n, \mathbf{x}_m)$ vertices, each of which is connected to the core vertex by a single edge. Outermost is *tier 2*: a layer of $\mathbf{x}_n$ vertices, each of which is connected to a vertex from tier 1 by a single edge. The local kernel at the core is set identically equal to zero, the local kernels at the $(\mathbf{x}_n, \mathbf{x}_m)$ and $\mathbf{x}_n$ vertices are set to $\alpha_{n,m}$ and $\alpha_n$ respectively. This tree satisfies all the three conditions C.1-C.3 (given in Section II-B) for it to be a junction tree for the MPF problem of ML decoding the STBC $\mathcal{C}$. Conditions C.1 and C.2 are easy to check. To illustrate the satisfiability of C.3 (the junction tree condition), Fig. 2 shows the subtree formed by the vertices whose local domains contain the symbol $\mathbf{x}_2$. Clearly this subtree is a connected graph.

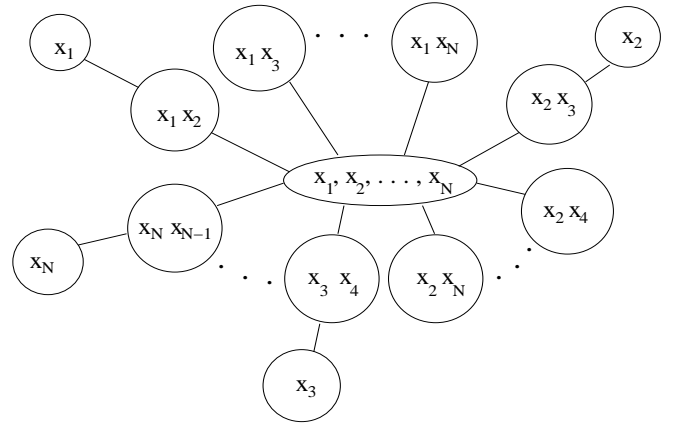

Fig. 1. A junction tree to decode an arbitrary STBC.

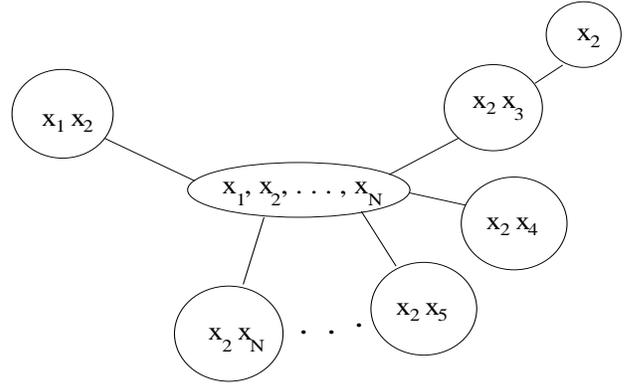

Fig. 2. Subtree formed by the vertices that contain $\mathbf{x}_2$.

## IV. FAST GDL DECODABLE SPACE-TIME BLOCK CODES

The junction tree of Fig. 1 has complexity order

$$\max_{v \in \mathcal{V}} |\mathcal{A}_{\mathcal{I}_v}| = |\mathcal{A}_{\mathcal{I}_{\{1,\ldots,N\}}}| = |\mathcal{C}|,$$

which is equal to the complexity order of brute-force ML decoding. There exist codes whose weight matrices $\{\mathbf{A}_i\}$ are such that the function $\alpha_{n,m}$ is identically equal to zero for all channel realizations $\mathbf{H}$ for certain pairs $(n, m)$. In such cases a number of 'factors' in the MPF formulation in (7) can be dropped, and this can lead to junction trees whose order of complexity is less than $|\mathcal{C}|$.

*Definition 1:* If an STBC $\mathcal{C}$ admits GDL decoding with complexity order less than $|\mathcal{C}|$ then we say that it is *fast GDL decodable*.

A number of properties of the GDL decoding of an STBC can be readily inferred from what are known as the *moral graph* of an STBC and the *core* of a junction tree. In the following subsection we introduce these notions, and in Section IV-B we give some results on the fast GDL decodability of STBCs based on these concepts.

### A. The Moral Graph and the Core

The local kernels $\alpha_{n,m}(\mathbf{x}_n, \mathbf{x}_m)$ arise from the cross terms $s_i s_j \xi_{i,j}$ (6), where $\xi_{i,j} = tr(\mathbf{H}(\mathbf{A}_i \mathbf{A}_j^H + \mathbf{A}_j \mathbf{A}_i^H)\mathbf{H}^H)$. It is well known [9]–[11] that a necessary and sufficient condition for $\xi_{i,j} = 0$ for any channel realization $\mathbf{H}$ is that $\mathbf{A}_i$ and $\mathbf{A}_j$ be Hurwitz-Radon orthogonal, i.e., $\mathbf{A}_i \mathbf{A}_j^H + \mathbf{A}_j \mathbf{A}_i^H = \mathbf{O}_{n_t}$. We say that two variables $\mathbf{x}_n$ and $\mathbf{x}_m$ *interfere* with each other if there exists a symbol $s_i$ in the encoding group $\mathbf{x}_n$ and a symbol $s_j$ in the encoding group $\mathbf{x}_m$ such that $\mathbf{A}_i \mathbf{A}_j^H + \mathbf{A}_j \mathbf{A}_i^H \neq \mathbf{O}_{n_t}$. If no such symbols $s_i, s_j$ exist we say that $\mathbf{x}_n$ and $\mathbf{x}_m$ are *non-interfering*. The local kernel $\alpha_{n,m}(\mathbf{x}_n, \mathbf{x}_m)$ is identically zero (and hence can be removed in the MPF formulation) for all channel realizations if and only if $\mathbf{x}_n$ and $\mathbf{x}_m$ are *non-interfering*. The moral graph [30] of the MPF formulation of ML decoding an STBC is a simple[2] graph whose vertices are the variables $\mathbf{x}_n$, $n = 1, \ldots, N$, and in which an edge exists between two vertices if and only if the two corresponding variables are interfering.

In the MPF formulation in (7) the kernels $\alpha_n(\mathbf{x}_n)$ arise from the terms $\xi_i s_i$ and $\xi_{i,i} s_i^2$. Recall that $\xi_{i,i} = ||\mathbf{H}\mathbf{A}_i||_F^2$ and hence is non-zero with probability 1. Thus, the kernels $\alpha_n$, $n = 1, \ldots, N$, are almost always non-zero and can not be removed from the MPF formulation. On the other hand, as we saw in the previous paragraph, some of the cross terms $\alpha_{n,m}$ can be made identically zero. This information about the cross terms is embedded in the moral graph of the code. Thus, all the information required to construct a junction tree for a code is contained in its moral graph. We now show how the problem of constructing a junction tree can be reduced to the construction of what we refer to as the *core*. Let $\mathcal{T}$ be a simple tree such that each vertex $v$ of $\mathcal{T}$ is associated with a variable list $\mathbf{x}_{\mathcal{I}_v}$ (for some $\mathcal{I}_v \subset \{1, \ldots, N\}$) and the kernel $\alpha_v(\mathbf{x}_{mathcalI_v}) = 0$.

*Definition 2:* The tree $\mathcal{T}$ is said to be a core for the STBC $\mathcal{C}$ if *(i)* it satisfies the junction tree condition (condition C.3 of Section II-B), and *(ii)* for every pair of neighboring vertices $(\mathbf{x}_n, \mathbf{x}_m)$ in the moral graph, there exists a vertex $v$ of $\mathcal{T}$ such that $\{\mathbf{x}_n, \mathbf{x}_m\} \subseteq \mathbf{x}_{\mathcal{I}_v}$.

Given a core $\mathcal{T}$, a junction tree for the STBC can be constructed as follows. For every pair $(\mathbf{x}_n, \mathbf{x}_m)$ of neighboring vertices in the moral graph, choose a vertex $v$ of $\mathcal{T}$ such that $\{\mathbf{x}_n, \mathbf{x}_m\} \subseteq \mathbf{x}_{\mathcal{I}_v}$. If $\mathcal{I}_v = \{n, m\}$ then set the local kernel at $v$ to $\alpha_{n,m}$, else attach a vertex $(\mathbf{x}_n, \mathbf{x}_m)$ with local kernel $\alpha_{n,m}$ to $v$ using a single edge. The set of $(\mathbf{x}_n, \mathbf{x}_m)$ vertices thus added to $\mathcal{T}$ form tier 1. Now, for each $n = 1, \ldots, N$, find a vertex of tier 1 that contains the variable $\mathbf{x}_n$ and attach the vertex $(\mathbf{x}_n)$ with the local kernel $\alpha_n$ to that vertex using a single edge. If there exists no tier 1 vertex that contains $\mathbf{x}_n$ then connect the $(\mathbf{x}_n)$ vertex with local kernel $\alpha_n$ to any vertex of tier 1 using a single edge. The set of $(\mathbf{x}_n)$ vertices thus added form tier 2. It is straightforward to show that the graph thus obtained is a junction tree for the STBC $\mathcal{C}$.

---

[2]A graph is said to be simple if it is undirected, unweighted with no loops or multiple edges.

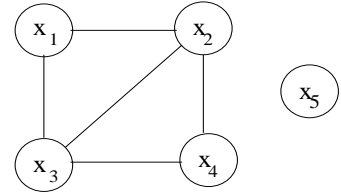

Fig. 3. Moral graph of Example 1.

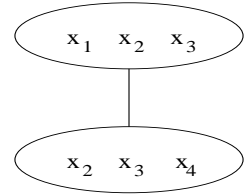

Fig. 4. The core of Example 1.

*Example 1:* Consider a code with $N = 5$ encoding groups and moral graph as shown in Fig. 3. There are five pairs of interfering symbols $\{(\mathbf{x}_1, \mathbf{x}_2), (\mathbf{x}_1, \mathbf{x}_3), (\mathbf{x}_2, \mathbf{x}_3), (\mathbf{x}_2, \mathbf{x}_4), (\mathbf{x}_3, \mathbf{x}_4)\}$. A core for this code is shown in Fig. 4. The core together with the tier 1 vertices is shown in Fig. 5. Note that the $(\mathbf{x}_2, \mathbf{x}_3)$ vertex of tier 1 could have been connected to the bottom vertex of the core as well. The complete junction tree is shown in Fig. 6. The vertex $(\mathbf{x}_5)$ has been connected to an arbitrarily chosen tier 1 vertex. The complexity order of this junction tree is $\max\{|\mathcal{A}_{\{1,2,3\}}|, |\mathcal{A}_{\{2,3,4\}}|\} < |\mathcal{C}|$, and hence this code is fast GDL decodable. ∎

Given the moral graph of an STBC, the problem of constructing a junction tree is equivalent to the problem of constructing a core. There is no unique core for a given STBC/moral graph, and different cores can lead to junction trees with different complexities. For instance, the graph with the single vertex $(\mathbf{x}_1, \mathbf{x}_2, \ldots, \mathbf{x}_N)$ can always be used as a core irrespective of the structure of the moral graph (see Fig. 1). However this would lead to junction trees with complexity order $|\mathcal{A}_{\{1,\ldots,N\}}| = |\mathcal{C}|$, which is equal to the order of brute-force ML decoding complexity.

When the moral graph is not edgeless, i.e., when there is at least one pair of interfering symbols, the complexity order of the junction tree is determined by the core vertices. Since every pair of interfering vertices must be contained within some 'larger' vertex of the core, the vertex $v$ of the junction tree with the largest $|\mathcal{A}_{\mathcal{I}_v}|$ belongs to the core. Thus, given an STBC/moral graph, *the problem of finding an efficient ML decoder is equivalent to one of constructing a core with the least complexity*.

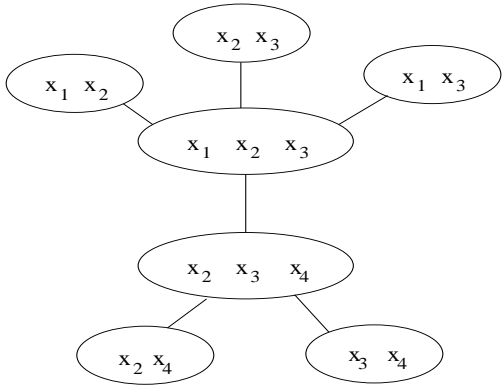

Fig. 5. The core $\mathcal{T}$ of Example 1 with tier 1 vertices.

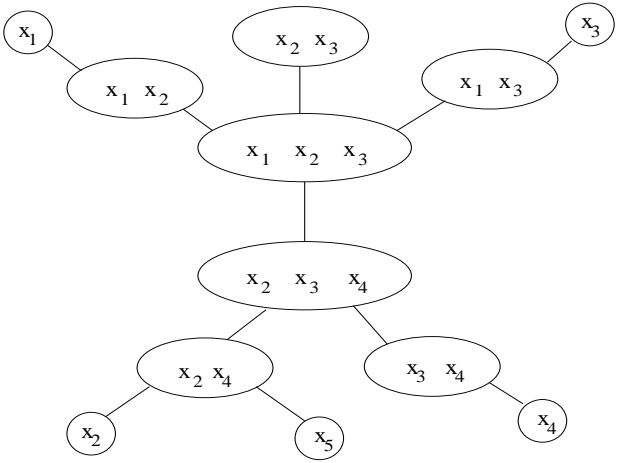

Fig. 6. The junction tree of Example 1.

When the moral graph is edgeless, i.e., when none of the symbols are interfering with each other, any tree $\mathcal{G}$ with $N$ vertices can be transformed into a junction tree by labeling the $N$ vertices with the local domains $(\mathbf{x}_n)$ and the local kernels $\alpha_n$, $n = 1, \ldots, N$ respectively. Since there are no cross terms $\alpha_{n,m}$ in the MPF formulation, the ML metric

$$f(s_1, \ldots, s_K) = \sum_{n=1}^{N} \alpha_n(\mathbf{x}_n) = \beta.$$

Since every variable $\mathbf{x}_n$ appears in exactly one of the vertices of $\mathcal{G}$, the tree $\mathcal{G}$ satisfies the junction tree condition as well. Hence $\mathcal{G}$ is a junction tree for the given STBC. The complexity order of this junction tree is $\max_{n=1}^{N} |\mathcal{A}_n| < |\mathcal{C}|$. Thus, STBCs with edgeless moral graphs are fast GDL decodable.

*Example 2:* All Orthogonal Designs [1] have edgeless moral graphs. For example, consider the Alamouti Code

$$\begin{bmatrix} s_1 + js_2 & -s_3 + js_4 \\ s_3 + js_4 & s_1 - js_2 \end{bmatrix},$$

where the real symbols $s_1, \ldots, s_4$ are encoded independently using a PAM constellation. This code has $N = 4$ encoding groups $\mathbf{x}_n = [s_n]$, $n = 1, \ldots, 4$. The moral graph, see Fig. 7, is edgeless. A junction tree for the Alamouti code is shown in Fig. 8. ∎

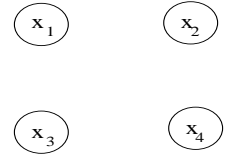

Fig. 7. Moral graph of the Alamouti Code.

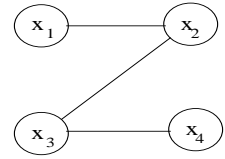

Fig. 8. A junction tree for the Alamouti Code.

### B. Fast GDL Decodable STBCs

We now give a sufficient condition for a code to admit fast GDL decoding.

*Lemma 2:* A code admits fast GDL decoding if its moral graph is not complete[3].

*Proof:* We prove the claim by constructing a core for such a code $\mathcal{C}$ with complexity order less than $|\mathcal{C}|$. Since the moral graph is not complete, there exist a pair of variables, say $\mathbf{x}_1$ and $\mathbf{x}_2$, that are not connected by an edge in the moral graph. Consider the tree shown in Fig. 9. There are $(N-1)$ variables in either of the vertices of this tree. It is straightforward to show that this tree satisfies both the conditions of Definition 2 to be a core for the given STBC. The order of GDL decoding complexity with this core is

$$\max\{|\mathcal{A}_{\{1,3,4,\ldots,N\}}|, |\mathcal{A}_{\{2,3,\ldots,N\}}|\} < |\mathcal{A}_{\{1,2,3,\ldots,N\}}| = |\mathcal{C}|,$$

and hence this code is fast GDL decodable. ∎

*Example T.2:* Continuing with Example T.1, the moral graph of the $2 \times 10$ Toeplitz code is given in Fig 10. The moral graph is not complete and hence this code admits fast GDL decoding. ∎

*Example G.2:* We now continue with Example G.1. First consider the naive choice of encoding groups with just two symbol groups. Since $\mathbf{A}_1 \mathbf{A}_5^H + \mathbf{A}_5 \mathbf{A}_1^H \neq \mathbf{O}_2$, the two symbol groups interfere and hence the moral graph is complete. Now consider the second choice of weight matrices and encoding groups with 8 symbol groups. The moral graph, shown in Fig. 11, is not complete and hence with this choice of weight matrices the Golden code admits fast GDL decoding. ∎

*Multigroup Decodable STBCs:* Let $\mathcal{G}$ be a junction tree for an STBC $\mathcal{C}$, and let there be $(g-1)$ edges $(u_k, v_k)$, $k = 1, \ldots, (g-1)$, of $\mathcal{G}$ such that $\mathcal{I}_{u_k} \cap \mathcal{I}_{v_k} = \phi$, the empty set. Let $\mathcal{G}_1, \ldots, \mathcal{G}_g$, be the $g$ disjoint subtrees of $\mathcal{G}$ obtained by removing these $(g-1)$ edges. Also, denote by $\mathbf{x}(\mathcal{G}_k)$ the union of the set of variables that appear in the local domains of $\mathcal{G}_k$.

*Theorem 1:* For $\mathcal{G}, \mathcal{G}_1, \ldots, \mathcal{G}_g$ described as above, we have:
1) $\mathbf{x}(\mathcal{G}_1), \ldots, \mathbf{x}(\mathcal{G}_g)$ is a partition of $\{\mathbf{x}_1, \ldots, \mathbf{x}_N\}$,

---
[3]A simple graph is said to be complete if every pair of distinct vertices is connected by an edge.

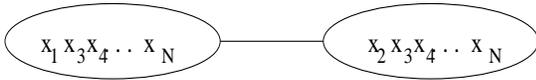

Fig. 9. The core used in the proof of Lemma 2.

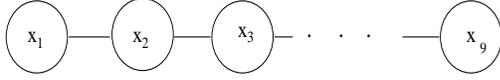

Fig. 10. Moral graph of the Toeplitz code in Example T.1.

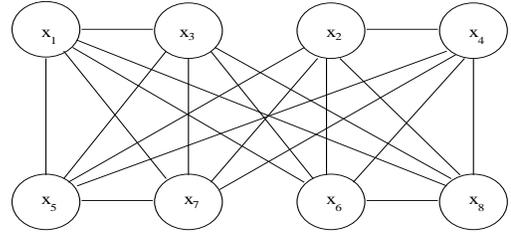

Fig. 11. Moral graph of the Golden Code.

2) for $k = 1, \ldots, g$, the tree $\mathcal{G}_k$ satisfies the junction tree condition, and
3) for each $k = 1, \ldots, g$, the ML solution of $\mathbf{x}(\mathcal{G}_k)$ can be obtained by running the GDL message-passing algorithm on $\mathcal{G}_k$.

*Proof:* The proof is given in Appendix A. ∎

We say that $\mathcal{G}_1, \ldots, \mathcal{G}_g$ is a *partition* of the junction tree $\mathcal{G}$, and that the STBC is GDL decodable using these $g$ independent junction trees. Each subtree $\mathcal{G}_k$ is composed only of a specific subset $\mathbf{x}(\mathcal{G}_k)$ of variables, hence for any vertex $v_k$ of $\mathcal{G}_k$ we have $\mathcal{I}_{v_k} \subsetneq \{1, \ldots, N\}$. Thus, the complexity order of $\mathcal{G}$ is

$$\max_{v \in \mathcal{G}} |\mathcal{A}_{\mathcal{I}_v}| = \max_{k \in \{1,\ldots,g\}} \max_{v_k \in \mathcal{G}_k} |\mathcal{A}_{\mathcal{I}_{v_k}}| < |\mathcal{C}|.$$

Thus, codes whose junction trees can be partitioned into two or more subtrees are fast GDL decodable.

*Example 3:* Consider the junction tree of Example 1 shown in Fig. 6. Among the 11 edges of this tree, the edge $(u, v)$ between the nodes $(\mathbf{x}_2, \mathbf{x}_4)$ and $(\mathbf{x}_5)$ is the only one such that $\mathcal{I}_u \cap \mathcal{I}_v = \phi$. Thus, in this case $g = 2$ and the two subtrees are shown in Fig. 12. The sets of variables $\mathbf{x}(\mathcal{G}_1) = \{\mathbf{x}_1, \mathbf{x}_2, \mathbf{x}_3, \mathbf{x}_4\}$ and $\mathbf{x}(\mathcal{G}_2) = \{\mathbf{x}_5\}$. The ML solutions of $\mathbf{x}(\mathcal{G}_1)$ and $\mathbf{x}(\mathcal{G}_2)$ can be obtained by running the GDL independently on $\mathcal{G}_1$ and $\mathcal{G}_2$ respectively. Note that the corresponding moral graph, shown in Fig. 3, is a disjoint union of $g = 2$ subgraphs. Further, the first subgraph is composed of variables from the set $\mathbf{x}(\mathcal{G}_1)$ and the second from the set $\mathbf{x}(\mathcal{G}_2)$. ∎

*Example 4:* All the three edges of the junction tree of the Alamouti code, shown in Fig. 8, satisfy the condition $\mathcal{I}_u \cap \mathcal{I}_v = \phi$. In this case $g = 4$, and the $k^{th}$ subtree $\mathcal{G}_k$ consists of a single vertex $(\mathbf{x}_k)$ with the local kernel $\alpha_k(\mathbf{x}_k)$. Note that the moral graph of this code, shown in Fig. 7, is disjoint union of $g = 4$ subgraphs, and the $k^{th}$ subgraph of the moral graph is composed of variables from $\mathbf{x}(\mathcal{G}_k)$. ∎

We will see in Lemmas 3 and 4 that the property of a junction tree to be partitioned into several smaller junction trees is related to *multigroup decodability* of a code, and as illustrated in the previous two examples, this property can be readily inferred from the moral graph. An STBC is said to be *multigroup* or $g$-group decodable [9]–[11] if $\{\mathbf{x}_1, \ldots, \mathbf{x}_N\}$ can be partitioned into $g$ subsets such that each subset of symbols can be ML decoded independently of other subsets. If the code generated by the $k^{th}$ group of symbols is $\mathcal{C}_k$, then the $k^{th}$ symbol group is ML decoded by the CML algorithm

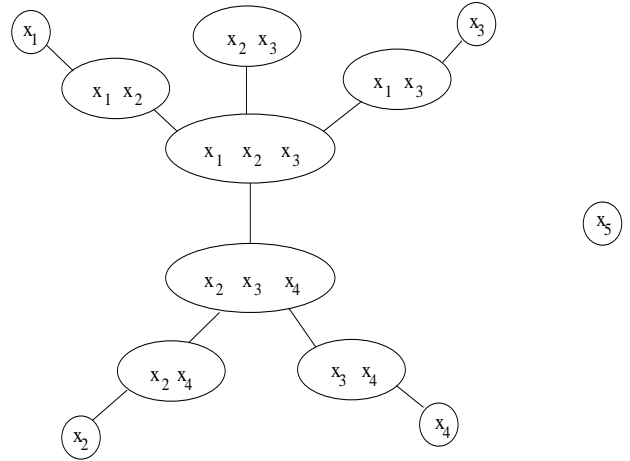

Fig. 12. The subtrees $\mathcal{G}_1$ and $\mathcal{G}_2$ of Example 3.

independent of other symbol groups as

$$\arg\min_{\mathbf{X}_k \in \mathcal{C}_k} ||\mathbf{Y} - \mathbf{H}\mathbf{X}_k||_F^2.$$

Thus, in order to decode $\mathcal{C}$, the $g$ subcodes $\mathcal{C}_1, \ldots, \mathcal{C}_g$ are decoded independently by the CML decoder. A necessary and sufficient condition for $g$-group decodability is that the weight matrices of the variables belonging to different subsets be Hurwitz-Radon orthogonal [9]–[11]. In terms of the GDL formulation, this translates to the variables belonging to different subsets being non-interfering.

*Lemma 3:* An STBC is $g$-group decodable if and only if its moral graph is a disjoint union of $g$ subgraphs.

*Proof:* The proof is straight forward. ∎

Using this lemma we see that any code with the moral graph of Fig. 3 is 2-group decodable, and that the Alamouti code is 4-group decodable.

*Lemma 4:* An STBC can be GDL decoded using a disjoint of union $g$ junction trees if and only if it is $g$-group decodable.

*Proof:* Suppose an STBC has a junction tree that can be be partitioned into $g$ subtrees $\mathcal{G}_1, \ldots, \mathcal{G}_g$. From Theorem 1, $\mathbf{x}(\mathcal{G}_1), \ldots, \mathbf{x}(\mathcal{G}_g)$ form a partition of the variables $\{\mathbf{x}_1, \ldots, \mathbf{x}_N\}$. Consider any two variables $\mathbf{x}_n$ and $\mathbf{x}_m$ belonging to distinct partitions. From Theorem 1, there exists no vertex in $\mathcal{G}$ whose local domain contains both $\mathbf{x}_n$ and $\mathbf{x}_m$. Thus, the global kernel does not involve the function $\alpha_{n,m}$, and hence $\mathbf{x}_n$ and $\mathbf{x}_m$ are non-interfering. We have thus shown that the variables belonging to the $g$ subsets $\mathbf{x}(\mathcal{G}_1), \ldots, \mathbf{x}(\mathcal{G}_g)$ are mutually non-interfering. Hence, the moral graph is a disjoint union of $g$-subgraphs, and from Lemma 3, the code is $g$-group decodable.

Suppose an STBC is $g$-group decodable. Then from Lemma 3, its moral graph is a disjoint union of $g$ subgraphs. For $k = 1, \ldots, g$, let $\Gamma_k \subset \{1, \ldots, N\}$ be the set of indices of the variables in the $k^{th}$ disjoint subgraph of the moral graph. One can then construct the $k^{th}$ disjoint subtree $\mathcal{G}_k$ of the junction tree $\mathcal{G}$ similar to the construction in Section III-B (see Fig. 1). The central node of $\mathcal{G}_k$ consists of all the variables $\mathbf{x}_n$, $n \in \Gamma_k$. The domains $(\mathbf{x}_n, \mathbf{x}_m)$ and $\mathbf{x}_n$, for $n, m \in \Gamma_k$ are then attached in two tiers, similar to the tree in Fig. 1. The junction tree $\mathcal{G}$ is obtained by arbitrarily connecting these $g$ subtrees using $(g-1)$ edges. It is straightforward to see that the resulting tree is a junction tree for the code, and that $\mathcal{G}_1, \ldots, \mathcal{G}_g$ form a partition of $\mathcal{G}$. Hence from Theorem 1, the code can be GDL decoded using a partition of $g$ disjoint junction trees. ∎

When a code is $g$-group decodable, the $k^{th}$ subcode is generated by the variables associated with the $k^{th}$ disjoint subgraph of the moral graph. A junction tree partition for this code can be obtained by constructing $g$ junction trees, one each for the $g$ subgraphs of the moral graph.

*Fast-Decodable STBCs:* An STBC is said to be *fast-decodable* [16] or *conditionally $g$-group decodable* [24] if there exists a subset $\Gamma \subsetneq \{1, \ldots, N\}$, such that the code generated by the variables $\mathbf{x}_n$, $n \in \Gamma$ is $g$-group decodable. The CML decoding algorithm to decode such a code proceeds as follows. For each of the $|\mathcal{A}_{\Gamma^c}|$ values that the variables $\mathbf{x}_{\Gamma^c}$ jointly assume, the conditionally optimal values of the remaining variables $\mathbf{x}_n$, $n \in \Gamma$ can be found out via $g$-group decoding. Note that each of these $g$ subcodes can themselves be fast-decodable (such codes are said to be *fast-group-decodable* [43]). From among these $|\mathcal{A}_{\Gamma^c}|$ values of $\mathbf{x}_{\{1,\ldots,N\}}$, the realization of $\mathbf{x}_{\{1,\ldots,N\}}$ that minimizes the ML metric $\|\mathbf{Y} - \mathbf{HX}\|_F^2$ is found out in a brute-force way. Let the $g$ subcodes correspond to the variables with index sets $\Gamma_1, \ldots, \Gamma_g$ and let the complexity order of decoding the $k^{th}$ subcode using CML be $\mathcal{O}_k$. For each $k = 1, \ldots, g$, the complexity order $\mathcal{O}_k \leq |\mathcal{A}_{\Gamma_k}|$. The complexity order of the CML algorithm is then

$$|\mathcal{A}_{\Gamma^c}| \max_{k \in \{1,\ldots,g\}} \mathcal{O}_k \leq |\mathcal{A}_{\Gamma^c}| \max_{k \in \{1,\ldots,g\}} |\mathcal{A}_{\Gamma_k}| < |\mathcal{C}|.$$

*Lemma 5:* An STBC is conditionally $g$-group decodable if and only if there exists a $\Gamma \subsetneq \{1, \ldots, N\}$ such that the moral graph of the reduced set of variables $\{\mathbf{x}_n | n \in \Gamma\}$ is a disjoint union of $g$ subgraphs.

*Proof:* Follows immediately from Lemma 3. ∎

From Lemmas 2 and 5 we see that conditionally $g$-group ML decodable codes admit fast GDL decoding.

*Example T.3:* Consider the Toeplitz code of Example T.2. With $\Gamma = \{1, \ldots, 9\} \setminus \{5\}$ we see that the moral graph generated by $\mathbf{x}_\Gamma$ is a disjoint union of 2 subgraphs (see Fig. 13). The first subgraph consists of the symbols $\mathbf{x}_1, \ldots, \mathbf{x}_4$ and the second subgraph consists of $\mathbf{x}_6, \ldots, \mathbf{x}_9$. Hence this code is conditionally 2-group decodable. Note that the code generated by the variables $\mathbf{x}_1, \ldots, \mathbf{x}_4$ is itself conditionally 2-group decodable where the two conditional groups are $\{\mathbf{x}_1\}$ and $\{\mathbf{x}_4\}$. Similarly the code generated by $\mathbf{x}_6, \ldots, \mathbf{x}_9$ is conditionally 2-group decodable as well. ∎

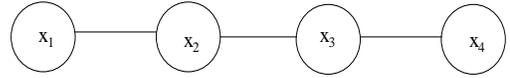

Fig. 13. Toeplitz code: Moral graph of the reduced set of variables $\mathbf{x}_\Gamma$.

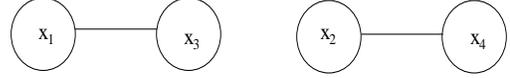

Fig. 14. Golden code: Moral graph of the reduced set of variables $\mathbf{x}_\Gamma$.

*Example G.3:* Consider the moral graph of the Golden code given in Fig. 11. For $\Gamma = \{1, 2, 3, 4\}$, the moral graph generated by the variables $\{\mathbf{x}_1, \ldots, \mathbf{x}_4\}$, shown in Fig 14, is a disjoint union of 2 subgraphs. The first subgraph consists of variables $\mathbf{x}_1, \mathbf{x}_3$ and the second subgraph consists of the variables $\mathbf{x}_2, \mathbf{x}_4$. Thus the Golden code is conditionally 2-group decodable. This fast-decodability property of the Golden code was first reported in [18], [29]. ∎

## V. GDL IS FASTER THAN CONDITIONAL ML DECODING

In this section we show that the number of computations involved in the GDL decoding of any STBC is less than that of CML decoding. As a first step towards this, we show that ML solutions can be obtained using only the single-vertex GDL algorithm followed by a 'traceback', rather than the more complex all-vertex GDL. This reduction is possible since we are only interested in the arg min of the objective functions at the various vertices, and not the objective functions themselves.

### A. Traceback

Let $\mathcal{G}$ be any junction tree for the STBC $\mathcal{C}$ with the encoding groups $\mathbf{x}_1, \ldots, \mathbf{x}_N$. We will now show that the ML solutions of $\{\mathbf{x}_n\}$ can be obtained by running the single-vertex GDL with any vertex $v_0$ as the root, followed by a traceback step. This is similar to the Viterbi's algorithm [32], where the actual ML metric of only the last state of the trellis is calculated and then the ML path is traced back to the first state.

Consider the single-vertex GDL message-passing schedule with $v_0$ as the root. Every vertex $u \neq v$ sends a message to its neighbor $p(u)$ on the unique path from $u$ to $v_0$, when it has received messages from all its other neighbors. While doing so it computes its partial state

$$\lambda_u(\mathbf{x}_{\mathcal{I}_u}) = \alpha_u(\mathbf{x}_{\mathcal{I}_u}) + \sum_{\substack{w \ adj \ u \\ w \neq p(u)}} \mu_{w,u}(\mathbf{x}_{\mathcal{I}_w \cap \mathcal{I}_u}),$$

and sends the message $\mu_{u,p(u)}$ as

$$\mu_{u,p(u)}(\mathbf{x}_{\mathcal{I}_u \cap \mathcal{I}_{p(u)}}) = \min_{\mathbf{x}_{\mathcal{I}_u \setminus \mathcal{I}_{p(u)}}} \lambda_u(\mathbf{x}_{\mathcal{I}_u}).$$

Note that this partial state $\lambda_u$ is different from the state $\sigma_u$ of $u$ at the end of the all-vertex GDL algorithm. These two

functions are related as

$$\sigma_u(\mathbf{x}_{\mathcal{I}_u}) = \lambda_u(\mathbf{x}_{\mathcal{I}_u}) + \mu_{p(u),u}(\mathbf{x}_{\mathcal{I}_{p(u)} \cap \mathcal{I}_u}),$$

where $\mu_{p(u),u}$ is the message from $p(u)$ to $u$ during the all-vertex GDL. However, the message $\mu_{p(u),u}$ is not generated during the single-vertex schedule. At the end of the single-vertex GDL, $v_0$ calculates its state $\sigma_{v_0}$, which is equal to the $\mathbf{x}_{\mathcal{I}_{v_0}}$-marginalization of $\beta$. The ML solution to $\mathbf{x}_{\mathcal{I}_{v_0}}$ is obtained as $\hat{\mathbf{x}}_{\mathcal{I}_{v_0}} = \arg\min \sigma_{v_0}(\mathbf{x}_{\mathcal{I}_{v_0}})$.

Let $u$ be any vertex such that the ML solution of the local domain of $p(u)$, i.e., $\hat{\mathbf{x}}_{\mathcal{I}_{p(u)}}$ is known. Partition $\mathbf{x}_{\mathcal{I}_u}$ into $\mathbf{x}_{A(u)} = \mathbf{x}_{\mathcal{I}_u \setminus \mathcal{I}_{p(u)}}$ and $\mathbf{x}_{B(u)} = \mathbf{x}_{\mathcal{I}_u \cap \mathcal{I}_{p(u)}}$. Note that both $\lambda_u$ and $\sigma_u$ are functions of both $\mathbf{x}_{A(u)}$ and $\mathbf{x}_{B(u)}$. Since the ML solution at $p(u)$ is known, the value $\hat{\mathbf{x}}_{B(u)}$ that minimizes $\sigma_u(\mathbf{x}_{A(u)}, \mathbf{x}_{B(u)})$ is known. Thus, the ML solution of $\mathbf{x}_{A(u)}$ is

$$\begin{aligned}\hat{\mathbf{x}}_{A(u)} &= \arg\min_{\mathbf{x}_{A(u)}} \sigma_u(\mathbf{x}_{A(u)}, \hat{\mathbf{x}}_{B(u)}) \\ &= \arg\min_{\mathbf{x}_{A(u)}} \lambda_u(\mathbf{x}_{A(u)}, \hat{\mathbf{x}}_{B(u)}) + \mu_{p(u),u}(\hat{\mathbf{x}}_{B(u)}) \\ &= \arg\min_{\mathbf{x}_{A(u)}} \lambda_u(\mathbf{x}_{A(u)}, \hat{\mathbf{x}}_{B(u)}).\end{aligned}$$

Hence, the ML solution at $u$ can be obtained merely from $\lambda_u$ and the ML solution at $p(u)$. This is possible since we are only interested in $\arg\min \sigma_u$ rather than $\sigma_u$ itself, and as shown above, $\arg\min \sigma_u$ can be obtained from $\lambda_u$ without calculating $\sigma_u$ explicitly. At the end of the single-vertex schedule, the solution at $v_0$ is first found, followed by all its neighbors, and then the neighbors of these vertices, and so on, until the ML solution of all the variables $\mathbf{x}_n$, $n = 1, \ldots, N$, are obtained. Since the all-vertex GDL is about four times as complex as the single-vertex GDL, this traceback algorithm provides a considerable reduction in complexity.

*Example 5:* The direction of messages for the single-vertex GDL problem on the subgraph $\mathcal{G}_1$ of Example 3 with root at the vertex $(\mathbf{x}_1, \mathbf{x}_3)$ is shown in Fig. 15. In this example, $p(b) = p(c) = a$, $p(d) = p(e) = p(g) = c$, $p(f) = e$, $p(h) = p(i) = g$, $p(u) = h$ and $p(v) = i$. At the end of the GDL schedule the state at the vertex $a$ is equal to the $(\mathbf{x}_1, \mathbf{x}_3)$-marginalization of the global kernel. The optimal $(\hat{\mathbf{x}}_1, \hat{\mathbf{x}}_3)$ is found out from $\sigma_a$ using $(|\mathcal{A}_1||\mathcal{A}_3| - 1)$ pairwise comparisons. Since $p(c) = a$, using the knowledge of $\hat{\mathbf{x}}_1, \hat{\mathbf{x}}_3$ and $\lambda_c$, the value of $\hat{\mathbf{x}}_2$ can then be found out. This step involves $(|\mathcal{A}_2| - 1)$ comparisons. Finally, given $\hat{\mathbf{x}}_2, \hat{\mathbf{x}}_3$ and $\lambda_g$ the value of $\hat{\mathbf{x}}_4$ can be obtained using $(|\mathcal{A}_4| - 1)$ comparisons. If $|\mathcal{A}_1| = \cdots = |\mathcal{A}_4| = q$, then finding the optimal $\mathbf{x}_n$, $n = 1, \ldots, 4$, using the single-vertex GDL and traceback involves $7q^3 + 4q^2 + 2q - 3$ operations. On the other hand, using the all-vertex GDL would cost $28q^3 + 12q^2 + 4q - 1$ operations. Comparing the leading order terms, we see that, traceback has enabled us to reduce the complexity by about 4 times. ∎

### B. GDL is faster than Conditional ML decoding

Before stating the results of this subsection, we define the GDL and conditional ML decoding complexities of an STBC, denoted by $\mathsf{C}_{\mathsf{GDL}}(\mathcal{C})$ and $\mathsf{C}_{\mathsf{CML}}(\mathcal{C})$ respectively. The GDL algorithm varies with the choice of the weight matrices, encoding groups and the junction tree. By $\mathsf{C}_{\mathsf{GDL}}(\mathcal{C})$ is meant the minimum among the complexities (the number of mathematical operations: multiplications, additions and comparisons) of all possible GDL algorithms that can be used to solve the ML decoding problem of $\mathcal{C}$. Similarly for the CML algorithm there can be more than one choice of reduced set of variables $\mathbf{x}_\Gamma$ which generate a multigroup decodable code. The complexity of conditional ML decoding then varies with this choice. By $\mathsf{C}_{\mathsf{CML}}(\mathcal{C})$ is meant the minimum among all possible conditional ML decoding complexities of code $\mathcal{C}$. By $\mathcal{O}_{\mathsf{GDL}}(\mathcal{C})$ and $\mathcal{O}_{\mathsf{CML}}(\mathcal{C})$ we denote the order of $\mathsf{C}_{\mathsf{GDL}}(\mathcal{C})$ and $\mathsf{C}_{\mathsf{CML}}(\mathcal{C})$ in terms of the signal set/constellation size.

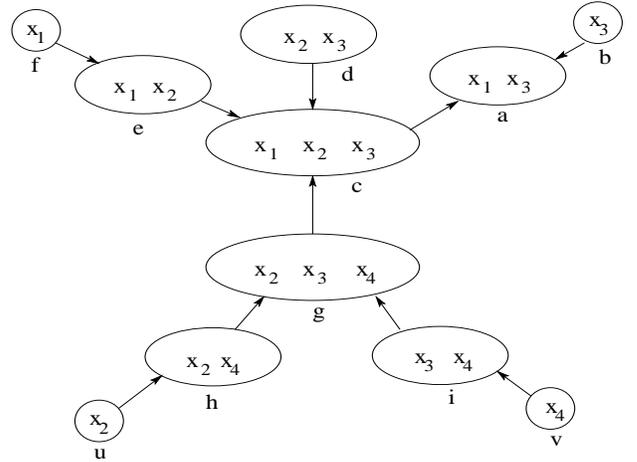

Fig. 15. Direction of messages for the single-vertex GDL for root vertex $a$.

*Order of decoding complexity:*

We now show that the order of GDL complexity of any code is upper bounded by the order its CML complexity.

*Theorem 2:* For any code $\mathcal{C}$, $\mathcal{O}_{\mathsf{GDL}}(\mathcal{C}) \leq \mathcal{O}_{\mathsf{CML}}(\mathcal{C})$.

*Proof:* Proof is given in Appendix B. ∎

The following example shows that there exist codes for which the GDL complexity order is strictly less. Thus the CML decoding algorithm is in general suboptimal in terms of reducing the ML decoding complexity.

*Example T.4:* The $2 \times 10$ Toeplitz code can be decoded using the junction tree given in Fig. 16 at the top of the next page. If the size of the complex HEX constellation used to encode the variables $\mathbf{x}_n = \begin{bmatrix} s_{2n-1} & s_{2n} \end{bmatrix}^T$ is $M$ then the complexity order of this junction tree is $|\mathcal{A}_{\{n,n-1\}}| = M^2$. The least complex CML algorithm proceeds as follows. The variables $\{\mathbf{x}_1, \ldots, \mathbf{x}_4\}$ and $\{\mathbf{x}_6, \ldots, \mathbf{x}_9\}$ are independently decoded after conditioning on $\mathbf{x}_5$. To decode $\{\mathbf{x}_1, \ldots, \mathbf{x}_4\}$, one first conditions on $\{\mathbf{x}_2, \mathbf{x}_3\}$ and finds the conditionally optimal values of $\mathbf{x}_1$ and $\mathbf{x}_4$ independently. The decoding of $\{\mathbf{x}_6, \ldots, \mathbf{x}_9\}$ proceeds in a similar way. Thus the CML complexity order is $M^4$. On the other hand, the brute-force decoding complexity, $|\mathcal{C}| = M^9$. Hence, for this code $\mathcal{O}_{\mathsf{GDL}} < \mathcal{O}_{\mathsf{CML}} < |\mathcal{C}|$. ∎

We now give two examples of families of STBCs for which $\mathcal{O}_{\mathsf{GDL}} < \mathcal{O}_{\mathsf{CML}}$.

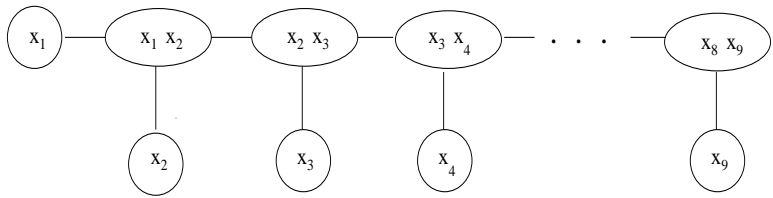

Fig. 16. Junction tree to decode the $2 \times 10$ Toeplitz code

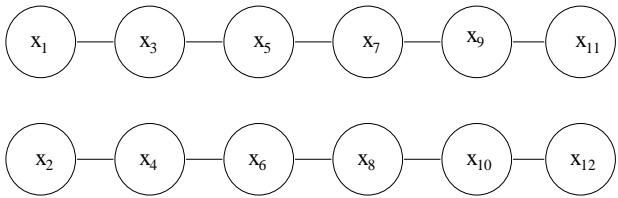

Fig. 17. The moral graph of the $4 \times 14$ OAC.

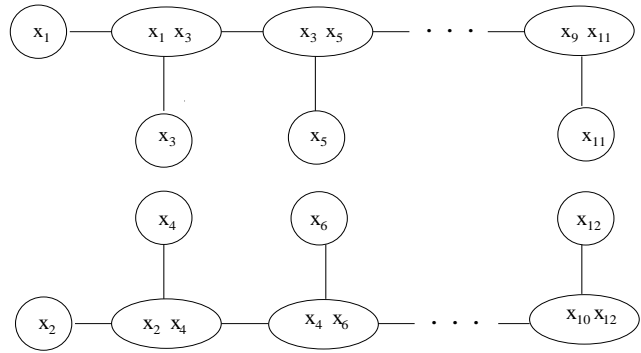

Fig. 18. A junction tree partition to decode the $4 \times 14$ OAC.

*1) Toeplitz Codes [39]:* Consider a $2 \times T$ Toeplitz Code, $T \geq 2$. This code consists of $K = 2(T-1)$ real symbols. We can construct a junction tree for this code similar to the one in Example T.4. The chain in this junction tree would extend till the $(\mathbf{x}_{T-2}, \mathbf{x}_{T-1})$ vertex. The complexity order of this junction tree is still $M^2$, irrespective of the value of $T$, where $M$ is the size of the complex constellation used to encode the symbols $\mathbf{x}_n$. The best ordering for conditional ML decoding this code is to first condition on the variable $\mathbf{x}_{\lfloor \frac{T-1}{2} \rfloor}$. This would result in two conditional ML decoding groups each of which generates a 'shorter' Toeplitz code whose delay is approximately $\frac{T}{2}$. Thus the CML decoding complexity grows with $M$ and $T$ as $M^{\log_2 T}$. It is interesting that though there is interference among the symbols, the GDL complexity is a constant independent of the number of symbols encoded by the code. These results can be extended to $n_t > 2$. For any $n_t \times T$ Toeplitz code there exists a junction tree whose complexity order is $M^{n_t}$. The CML decoding complexity however grows with the delay $T$.

*2) Overlapped Alamouti Codes (OACs) [40]:* These codes are 2-group ML decodable and are available for all choices of $T \geq n_t \geq 2$. They can be GDL decoded with complexity order $M^{\lfloor \frac{n_t+1}{2} \rfloor}$. The CML decoding complexity on the other hand grows with the number of symbols or equivalently with the delay $T$. For example, for $n_t = 4$, the CML complexity grows as $M^{\lceil \log_2(\frac{T}{2}) \rceil}$. As an example we construct a junction tree for the $4 \times 14$ OAC and show that its complexity order less than the CML decoding complexity.

The $4 \times 14$ OAC consists of 24 real symbols $s_1, \ldots, s_{24}$. Define the auxiliary variables $z_1, \ldots, z_{12}$ as $z_n = s_{2n-1} + j s_{2n}$. The design in terms of these auxiliary variables is given in (9) at the top of the next page. The variables $z_n$, $n = 1, \ldots, 12$, are encoded independently using a complex constellation of size $M$. Choose the encoding groups as $\mathbf{x}_n = \begin{bmatrix} s_{2n-1} & s_{2n} \end{bmatrix}^T$ for $n = 1, \ldots, 12$. The moral graph for the code is given in Fig. 17. The moral graph is not complete and hence from Lemma 2, this code admits fast GDL decoding. Since the moral graph is a disjoint union of two subgraphs, from Lemma 3, this code is 2-group decodable. A junction tree partition to decode this code is shown in Fig. 18. Note that this partition consists of 2 subtrees, each of which is a junction tree for the subcode generated by the 2 ML decoding groups. The complexity order of this junction tree partition is $M^2$. When CML decoding is used, the least achievable complexity order is $M^3$. We explain the CML decoding for the first ML decoding group. The decoding of the second group is similar. On fixing the value of $\mathbf{x}_5$, we get two conditional decoding groups. The first group $\{\mathbf{x}_1, \mathbf{x}_3\}$ is jointly decoded with complexity $M^2$ for each value of $\mathbf{x}_5$. The second group, $\{\mathbf{x}_7, \mathbf{x}_9, \mathbf{x}_{11}\}$, is again conditionally 2-group decoded with the two conditional groups being $\{\mathbf{x}_7\}$ and $\{\mathbf{x}_{11}\}$.

*Exact decoding complexity:*

Almost all STBCs of interest have the property that each encoding group has the same number of real symbols, say $t$, and the signal set size of all the groups are equal, i.e., $|\mathcal{A}_1| = |\mathcal{A}_2| = \cdots = |\mathcal{A}_N|$. If the average number of information bits carried by each real symbol is $\log_2 q$ then the signal set size $|\mathcal{A}_n| = q^t$. For example, when $t = 2$ the real symbols $\{s_i\}$ are encoded pairwise, and $q^2$ is the size of the complex constellation used to encode each $\mathbf{x}_n$. For the sake of analytical tractability, and considering the widespread prevalence STBCs of this type in the literature, we restrict our analysis of the exact GDL and CML complexities to codes wherein the number of real symbols in each encoding group is the same and $|\mathcal{A}_n| = q^t$.

Let $\mathcal{C}$ be any code where all the symbols $\mathbf{x}_n$, $n = 1, \ldots, N$, are mutually interfering. We will refer to such codes as being *fully-interfering*. In Appendix C we compute the exact CML and GDL complexities of such a fully-interfering STBC. The

$$\mathbf{S} = \begin{bmatrix} z_1 & 0 & z_3 & -z_2^* & z_5 & -z_4^* & z_7 & -z_6^* & z_9 & -z_8^* & z_{11} & -z_{10}^* & 0 & -z_{12}^* \\ 0 & z_1^* & z_2 & z_3^* & z_4 & z_5^* & z_6 & z_7^* & z_8 & z_9^* & z_{10} & z_{11}^* & z_{12} & 0 \\ 0 & -z_2^* & z_1 & -z_4^* & z_3 & -z_6^* & z_5 & -z_8^* & z_7 & -z_{10}^* & z_9 & -z_{12}^* & z_{11} & 0 \\ z_2 & 0 & z_4 & z_1^* & z_6 & z_3^* & z_8 & z_5^* & z_{10} & z_7^* & z_{12} & z_9^* & 0 & z_{11}^* \end{bmatrix}. \quad (9)$$

CML algorithm performs a brute-force minimization of the ML metric over all $q^{Nt}$ values of $(s_1, \ldots, s_{Nt})$. Its complexity is

$$\mathsf{C}_{\mathsf{CML}}(\mathcal{C}) = q^{Nt}\left(3\binom{Nt}{2} + 5Nt\right) - 1. \quad (10)$$

To GDL decode this STBC, we use the junction tree of Fig. 1 in Section III-B. We employ a single-vertex GDL schedule with the root at any one of the $(\mathbf{x}_n, \mathbf{x}_m)$ vertices followed by traceback (using the core vertex as the root will contribute to the leading order term $q^{Nt}$, which is avoided here). The complexity of this GDL decoder is given in (11) at the top of the next page. Comparing the leading terms of (10) and (11), we see that when the real symbols $\{s_i\}$ are encoded independently of each other i.e., when $t = 1$, the GDL is about 3 times less complex as the CML. When the symbols are encoded pairwise using a complex constellation, i.e., when $t = 2$, the GDL is approximately 12 times less complex than the CML decoder. For example, for any STBC obtained from Cyclic Division Algebras [38] that is not multigroup or conditionally multigroup decodable, the GDL decoder gives roughly a 12 times reduction in complexity compared to the CML decoder.

*Example 6:* Consider the following 2 antenna code obtained from a Cyclic Division Algebra [38]

$$\begin{bmatrix} s_1 + js_2 + \gamma(s_3 + js_4) & \delta\left(s_5 + js_6 - \gamma(s_7 + js_8)\right) \\ s_5 + js_6 + \gamma(s_7 + js_8) & s_1 + js_2 - \gamma(s_3 + js_4) \end{bmatrix},$$

where $\gamma = e^{j\frac{2\pi}{8}}$, and $\delta$ is any complex number which is transcendental over the field $\mathbb{Q}(\sqrt{\gamma})$. The complex symbols $s_{2n-1} + js_{2n}$, $n = 1, \ldots, 4$, are encoded using the 8-PSK signal set. For this code, there are $N = 4$ encoding groups, $\mathbf{x}_n = \begin{bmatrix} s_{2n-1} & s_{2n} \end{bmatrix}^T$ for $n = 1, \ldots, 4$, $t = 2$ and $q = \sqrt{8}$. All the four symbol groups are mutually interfering, and hence this STBC is fully-interfering. From (10), the CML decoder for this code involves $507,903$ mathematical operations. On the other hand, using (11), we see that the GDL decoder involves only $26,718$ operations, which is about 19 times less than the CML complexity. ∎

*Example 7:* Consider the following Field Extension code [38] for $n_t = 3$ transmit antennas

$$\begin{bmatrix} s_1 + js_2 & \gamma(s_5 + js_6) & \gamma(s_3 + js_4) \\ s_3 + js_4 & s_1 + js_2 & \gamma(s_5 + js_6) \\ s_5 + js_6 & s_3 + js_4 & s_1 + js_2 \end{bmatrix},$$

where $\gamma = e^{j\frac{2\pi}{6}}$ and the complex symbols $s_{2n-1} + js_{2n}$, $n = 1, \ldots, 3$ are encoded using the 8-PSK signal set. This code has $N = 3$ encoding groups, $\mathbf{x}_n = \begin{bmatrix} s_{2n-1} & s_{2n} \end{bmatrix}^T$ for $n = 1, \ldots, 3$, $t = 2$ and $q = \sqrt{8}$. This STBC is fully-interfering, and the CML and the GDL decoders for this code involve $38,399$ and $2,758$ operations respectively. Thus the GDL decoder provides a complexity reduction of the factor of 14 compared to the CML decoder. ∎

The number of computations involved in the GDL decoder is less than that of the CML decoder not just for fully-interfering codes, but for any STBC.

*Theorem 3:* Let $\mathcal{C}$ be any STBC such that the number of real symbols per each encoding group of $\mathcal{C}$ is same, and the signal set size for each of the encoding groups is equal. Then $\mathsf{C}_{\mathsf{GDL}}(\mathcal{C}) < \mathsf{C}_{\mathsf{CML}}(\mathcal{C})$.

*Proof:* Proof is given in Appendix D. ∎

From Theorem 2 and Example T.4, we see that the GDL algorithm can provide improvements over CML decoders in terms of the order of ML decoding complexity as well.

### C. Reduction in complexity with PAM signal sets

When a real symbol is encoded using a PAM signal set, the optimal value of that variable, conditioned on the values of other information symbols, can be found by scaling and hard-limiting. This technique has been widely used in the literature [18], [20], [26], [29], and can lead to gains in the order of the CML decoding complexity. In this subsection we show that such a reduction in complexity is possible with GDL as well.

We will now describe how a variable $\mathbf{x}_{n_0}$, $n_0 \in \{1, \ldots, N\}$, (not necessarily a PAM encoded single real symbol) can be *removed* from the GDL formulation. The global metric $\beta$ can be split into terms involving $\mathbf{x}_{n_0}$ and terms not involving $\mathbf{x}_{n_0}$ as

$$\beta = \alpha_{n_0}(\mathbf{x}_{n_0}) + \sum_{m \in \mathcal{N}(n_0)} \alpha_{n_0,m}(\mathbf{x}_{n_0}, \mathbf{x}_m)$$
$$+ \sum_{n \neq n_0} \alpha_n(\mathbf{x}_n) + \sum_{\substack{n < m \\ n,m \neq n_0}} \alpha_{n,m}(\mathbf{x}_n, \mathbf{x}_m),$$

where $\mathcal{N}(n_0)$ is the set of indices of those variables that are neighbors of $\mathbf{x}_{n_0}$ in the moral graph of the code. Define the functions

$$h_{n_0}(\mathbf{x}_{\mathcal{N}(n_0)}) = \min_{\mathbf{x}_{n_0}} \alpha_{n_0}(\mathbf{x}_{n_0}) + \sum_{m \in \mathcal{N}(n_0)} \alpha_{n_0,m}(\mathbf{x}_{n_0}, \mathbf{x}_m),$$

$$\beta'(\mathbf{x}_{\{n_0\}^c}) = \min_{\mathbf{x}_{n_0}} \beta(\mathbf{x}_1, \ldots, \mathbf{x}_N).$$

Then we have $\beta'(\mathbf{x}_{\{n_0\}^c}) =$

$$h_{n_0}(\mathbf{x}_{\mathcal{N}(n_0)}) + \sum_{n \neq n_0} \alpha_n(\mathbf{x}_n) + \sum_{\substack{n < m \\ n,m \neq n_0}} \alpha_{n,m}(\mathbf{x}_n, \mathbf{x}_m),$$

and the ML solution for $\mathbf{x}_n$, $n \neq n_0$,

$$\hat{\mathbf{x}}_{\{n_0\}^c} = \arg\min \beta'(\mathbf{x}_{\{n_0\}^c}).$$

Given the function $h_{n_0}(\mathbf{x}_{\mathcal{N}(n_0)})$, the ML decoding of $\mathcal{C}$ is equivalent to minimizing $\beta'$. This minimization can be solved

$$\mathsf{C}_{\mathsf{GDL}}(\mathcal{C}) = q^{Nt}\binom{N}{2} + q^{(N-2)t} + q^{2t}\left[\binom{N}{2}(2t-1) + N + 1\right] + q^{t}\left[\binom{N}{2}(2t^2 - t) + N(t^2 + 3t)\right] - 2. \quad (11)$$

using the GDL. If the function $h_{n_0}$ can be computed with sufficiently low complexity, using $\beta'$ rather than $\beta$ to ML decode $\mathcal{C}$ can lead to gains in the decoding complexity.

As we show now, when $\mathbf{x}_{n_0}$ is a $q$-ary PAM encoded single real symbol, $h_{n_0}$ can be computed with reduced complexity using scaling and hard-limiting. For each $\mathbf{x}_{\mathcal{N}(n_0)} \in \mathcal{A}_{n_0}$,

$$h_{n_0} = \min_{\mathbf{x}_{n_0}} \left[ \xi_{n_0,n_0}\mathbf{x}_{n_0}^2 + \left( \xi_{n_0} + \sum_{\substack{m \in \mathcal{N}(n_0) \\ i \in \psi(m)}} \xi_{n_0,i}s_i \right)\mathbf{x}_{n_0} \right]$$
$$= \min_{\mathbf{x}_{n_0}} \xi_{n_0,n_0}\left[ \left( \mathbf{x}_{n_0} - \frac{\zeta}{2\xi_{n_0,n_0}} \right)^2 - \frac{\zeta^2}{4\xi_{n_0,n_0}^2} \right],$$

where $\zeta = \xi_{n_0} + \sum_{m \in \mathcal{N}(n_0)} \sum_{i \in \psi(m)} \xi_{n_0,i}s_i$. The optimal value $\hat{\mathbf{x}}_{n_0}$ that minimizes $h_{n_0}$ for a given value of $\mathbf{x}_{\mathcal{N}(n_0)}$ can be found by the scaling and hard-limiting step given in (12) at the top of the next page, where $rnd(\cdot)$ is the nearest integer function. This step has a constant complexity independent of $q$. The value of $h_{n_0}$ can then be calculated as

$$h_{n_0}(\mathbf{x}_{\mathcal{N}(n_0)}) = \xi_{n_0,n_0}\left[ \left( \hat{\mathbf{x}}_{n_0} - \frac{\zeta}{2\xi_{n_0,n_0}} \right)^2 - \frac{\zeta^2}{4\xi_{n_0,n_0}^2} \right]. \quad (13)$$

We now use GDL to compute $h_{n_0}$ itself. From (13), we see that the function $h_{n_0}$ depends on $\mathbf{x}_{\mathcal{N}(n_0)}$ only through

$$\zeta = \xi_{n_0,n_0} + \sum_{j=1}^{p} \omega_{m_j}(\mathbf{x}_{m_j}), \text{ where,}$$

$\mathcal{N} = \{m_1, \ldots, m_p\}$ and $\omega_{m_j}(\mathbf{x}_{m_j}) = \sum_{i \in \psi(m_j)} \xi_{n_0,i}s_i$. Now consider the junction tree for this problem shown in Fig. 19, where the local kernel at the central vertex is $\xi_{n_0,n_0}$, and the local kernel at the vertex $(\mathbf{x}_{m_j})$ is $\omega_{m_j}$. It is straightforward to show that $\zeta$ is equal to the state of the central vertex of Fig. 19 at the end of the single-vertex GDL schedule rooted at this node. Using the table of values of $\zeta$ thus obtained, one can then compute $h_{n_0}$ using (12) and (13). Thus, the function $h_{n_0}$ can be computed with order of complexity $|\mathcal{A}_{\mathcal{N}(n_0)}|$ instead of the brute-force complexity order $q|\mathcal{A}_{\mathcal{N}(n_0)}|$.

If $\mathcal{G} = (\mathcal{V}, \mathcal{E})$ is a junction tree for $\beta$, and $\mathcal{G}'(\mathcal{V}', \mathcal{E}')$ is a junction tree for $\beta'$, such that

$$\max_{v' \in \mathcal{G}'} |\mathcal{A}_{\mathcal{I}_{v'}}| < \max_{v \in \mathcal{G}} |\mathcal{A}_{\mathcal{I}_v}| \text{ and}$$
$$|\mathcal{A}_{\mathcal{N}(n_0)}| < \max_{v \in \mathcal{G}} |\mathcal{A}_{\mathcal{I}_v}|,$$

then ML decoding the code using the junction tree $\mathcal{G}'$ provides an improvement in the complexity order compared to using the junction tree $\mathcal{G}$.

*Lemma 6:* If the core $\mathcal{T}$ of $\mathcal{G}$ has only one vertex containing the variable $\mathbf{x}_{n_0}$, then the tree $\mathcal{T}'$ obtained by removing $\mathbf{x}_{n_0}$ from this vertex of $\mathcal{T}$ is a core for the GDL minimization of $\beta'$.

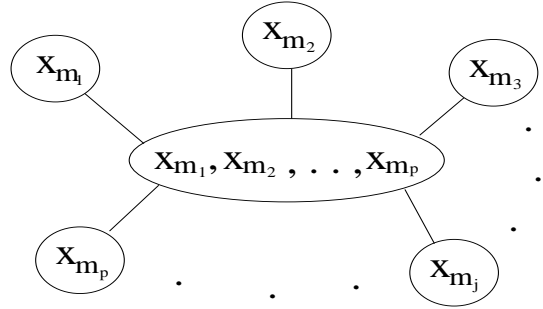

Fig. 19. A junction tree to compute $\zeta$.

*Proof:* We will show that $\mathcal{T}'$ satisfies both the conditions of Definition 2 for minimizing $\beta'$. Since $\mathcal{T}$ satisfies the junction tree condition for all the variables $\mathbf{x}_n$, $n = 1, \ldots, N$, the tree $\mathcal{T}'$, obtained by removing the only occurrence of $\mathbf{x}_{n_0}$, satisfies the junction tree condition for $\mathbf{x}_{n_0}$, $n \neq n_0$. For every $n, m \neq n_0$ there exists a $v \in \mathcal{V}$ such that $\{n, m\} \subseteq \mathcal{I}_v$, and hence there exists a $v' \in \mathcal{V}'$ such that $\{n, m\} \subseteq \mathcal{I}_{v'}$. Suppose $v_0 \in \mathcal{V}$ is the only vertex of $\mathcal{G}$ that contains $\mathbf{x}_{n_0}$. Because $\mathcal{T}$ is a core for the minimization of $\beta$, $\mathcal{N}(n_0) \subseteq \mathcal{I}_{v_0}$ and hence, this vertex in $\mathcal{T}'$ contains the argument of $h_{n_0}$ as a subset of its local domain. Therefore, $\mathcal{T}'$ can be used as a core for minimizing $beta'$. ∎

This technique of removing a PAM encoded variable can be generalized to any set $\mathcal{R} \subseteq \{1, \ldots, N\}$ of variables that satisfies the condition given in Lemma 7 below. In this case, the variables $\mathbf{x}_n$, $n \in \mathcal{R}$, are removed one by one from the GDL formulation, in an arbitrary order, using the same technique as above.

*Lemma 7:* The PAM encoded set of variables $\mathbf{x}_{\mathcal{R}}$ can be removed from the GDL formulation using scaling and hard-limiting if and only if the subgraph of the moral graph generated by these variables is edgeless.

*Proof:* Let $\mathcal{R} = \{n_1, \ldots, n_{|\mathcal{R}|}\}$, and let the chosen order of removal be $n_1, n_2, \ldots, n_{|\mathcal{R}|}$. The variable $\mathbf{x}_{n_1}$ can be removed using the technique described in this subsection, irrespective of the choice of $n_2, \ldots, n_{|\mathcal{R}|}$. Suppose there exists an $n_r \in \mathcal{R}$, such that $n_r \in \mathcal{N}(n_1)$. Then, while removing $\mathbf{x}_{n_r}$, one is faced with the minimization of the function

$$h_{n_1}(\mathbf{x}_{\mathcal{N}(n_1)}) + \alpha_{n_r}(\mathbf{x}_{n_r}) + \sum_{m \in \mathcal{N}(n_r)} \alpha_{n_r,m}(\mathbf{x}_{n_0}, \mathbf{x}_m)$$

over the variable $\mathbf{x}_{n_r}$. However, $h_{n_1}$ is not a quadratic function of $\mathbf{x}_{n_r}$, and hence minimization of the above expression via completion of squares, scaling and hard-limiting is not possible. On the other hand, when $n_r \notin \mathcal{N}(n_1)$, this step of minimizing $h_{n_1}$ does not arise during the removal of $\mathbf{x}_{n_r}$ from the GDL formulation, and hence $\mathbf{x}_{n_r}$ can be removed using scaling and hard-limiting. ∎

For example, when a conditionally $g$-group decodable code is to be decoded, one PAM encoded symbol from each of the

$$\hat{\mathbf{x}}_{n_0} = \min\left\{\max\left\{rnd\left(\frac{q-1}{2} - \frac{\zeta}{2\xi_{n_0,n_0}}\right), 0\right\}, q-1\right\} - \frac{q-1}{2}. \tag{12}$$

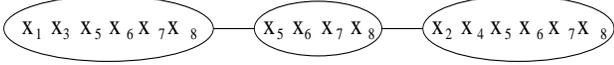

Fig. 20. A junction tree core $\mathcal{T}$ to decode the Golden Code.

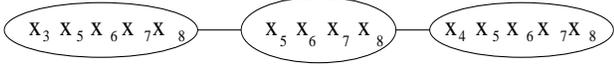

Fig. 21. A junction tree core $\mathcal{T}'$ for the Golden code that exploits the structure of PAM signal set.

$g$ conditional groups can be removed via scaling and hard-limiting.

*Example G.4:* Consider the junction tree core $\mathcal{T}$ for the Golden code shown in Fig. 20. From Lemma 7 and the moral graph of the Golden code given in Fig. 11, we see that the variables $\mathbf{x}_1$ and $\mathbf{x}_2$ can be removed using scaling and hard-limiting. Using Lemma 6 we get the junction tree core $\mathcal{T}' = (\mathcal{V}', \mathcal{E}')$ shown in Fig. 21. Since $|\mathcal{N}_1| = |\mathcal{N}_2| = 5$, the functions $h_1$ and $h_2$ can be computed with complexity order $q^5$, where $q$ is the size of the PAM signal set used to encode the information symbols. Also, $\max_{v' \in \mathcal{V}'} |\mathcal{A}_{\mathcal{I}_{v'}}| = q^5$, and hence the single-vertex GDL schedule and traceback can be implemented with order of complexity $q^5$. Hence, the order of complexity for GDL decoding of the Golden code using $\mathcal{T}'$ is $q^5$, whereas the complexity order of using $\mathcal{T}$ is $q^6$. The removal of the variables $\mathbf{x}_1$ and $\mathbf{x}_2$ has enabled the reduction of the GDL complexity order from $q^6$ to $q^5$. The total number of mathematical operations involved in the GDL decoding of the Golden code using $\mathcal{T}'$ is $42q^5 + 6q^4 + 21q^2 + 52q - 5$. The CML decoder [18], [29], on the other hand, involves $76q^5 + 43q^4 - 1$ operations. Comparing the leading order terms, we see that the GDL decoder is about 1.8 times as fast as the CML decoder. For instance, when $q = 2$ or $4$ (corresponding to the rates 4 and 8 bits per channel use), the GDL decoder gives a complexity reduction of 1.9 compared to the CML decoding algorithm.

On the other hand, consider the naive choice of symbol groups

$$\mathbf{x}_1 = \begin{bmatrix} s_1 & s_2 & s_3 & s_4 \end{bmatrix}^T, \ \mathbf{x}_2 = \begin{bmatrix} s_5 & s_6 & s_7 & s_8 \end{bmatrix}^T,$$

given in Example G.1. The signal set size for each of these two symbol groups is $q^4$. Since the two symbol groups are interfering, any choice of junction tree $\tilde{\mathcal{G}} = (\tilde{\mathcal{V}}, \tilde{\mathcal{E}})$ must involve a vertex $v_0$ that contains both the variables $\mathbf{x}_1, \mathbf{x}_2$. The GDL single-vertex decoding complexity has the complexity order $\max_{\tilde{v} \in \tilde{\mathcal{V}}} q^{4|\mathcal{I}_{\tilde{v}}|} \geq q^8$, which is equal to the order of brute-force ML decoding complexity. ∎

## VI. CONCLUSION

The CML decoding algorithm minimizes the ML metric $\beta(\mathbf{x}_1, \ldots, \mathbf{x}_N)$ via removing a subset of variables from the problem formulation by minimizing $\beta$ for each instantiation of this subset of variables. This subset of variables is chosen in such a way that the reduced problem, obtained after their removal from $\beta$, splits into multiple, independent, less complex minimization problems. The GDL, on the other hand, computes various partial sums and marginalizations of $\beta$ involving the 'smaller', less complex functions $\alpha_n$, $\alpha_{n,m}$, and utilizes these intermediate functions to efficiently arrive at the ML solution. In this paper, we have introduced this GDL based ML decoding framework, and shown that the GDL decoder is superior to the CML decoder in terms of complexity. The results of this paper have brought to light the following relevant problems that need to be addressed.

- Proving the optimality or otherwise of GDL based decoders in minimizing the complexity of ML decoding an STBC.
- Given an STBC $\mathcal{C}$, finding the optimal choice of weight matrices, encoding groups and signal sets, which will minimize the GDL decoding complexity of the code.
- Constructing codes with better rate-decoding complexity tradeoff than that of the known codes using the GDL decoders.
- Both GDL and CML decoding algorithms depend on the Hurwitz-Radon orthogonality of weight matrices to obtain low complexity ML decoders. Is there any other algebraic property of a code that can be exploited to design low complexity ML decoders? Can it lead to further improvement in the rate-decoding complexity tradeoff?

ACKNOWLEDGMENT

This work was supported partly by the DRDO-IISc program on Advanced Research in Mathematical Engineering through a research grant, and partly by the INAE Chair Professorship grant to B. S. Rajan. The authors thank K. Pavan Srinath for useful discussions on this subject.

## APPENDIX A
## PROOF OF THEOREM 1

First we will show that $\mathbf{x}(\mathcal{G}_1), \ldots, \mathbf{x}(\mathcal{G}_g)$ is a partition of $\{\mathbf{x}_1, \ldots, \mathbf{x}_N\}$. It is clear that $\cup_{k=1}^g \mathbf{x}(\mathcal{G}_k) = \{\mathbf{x}_1, \ldots, \mathbf{x}_N\}$. Enough to show that for any $\ell \neq k$, $\mathbf{x}(\mathcal{G}_\ell) \cap \mathbf{x}(\mathcal{G}_k) = \phi$. Suppose this is not true. There exists a variable $\mathbf{x}_n$ that appears in the local domains of at least one of the vertices in each of $\mathcal{G}_\ell$ and $\mathcal{G}_k$. Since $\mathcal{G}$ satisfies the junction tree condition, the local domains of all the vertices on the unique path between these two vertices in $\mathcal{G}$ contain the variable $\mathbf{x}_n$. Further, this unique path contains at least one of the edges $(u_k, v_k)$, $k = 1, \ldots, (g-1)$. Thus, there exists a $k$ such that $\mathcal{I}_{u_k} \cap \mathcal{I}_{v_k} \supseteq \{n\}$, and hence $\mathcal{I}_{u_k} \cap \mathcal{I}_{v_k} \neq \phi$, a contradiction. Thus $\mathbf{x}(\mathcal{G}_1), \ldots, \mathbf{x}(\mathcal{G}_g)$ is a partition of $\{\mathbf{x}_1, \ldots, \mathbf{x}_N\}$.

We will now show that for each $k = 1, \ldots, g$, the tree $\mathcal{G}_k$ satisfies the junction tree condition. Let $\mathbf{x}_n$ be any variable from the set $\mathbf{x}(\mathcal{G}_k)$. From the first result of this theorem, $\mathbf{x}_n$ appears in the local domains of the vertices of $\mathcal{G}_k$ only. Thus the subgraph of $\mathcal{G}$ formed by vertices containing $\mathbf{x}_n$ is a subgraph of $\mathcal{G}_k$. Since $\mathcal{G}$ satisfies the junction tree condition, this subgraph is a connected graph. Hence $\mathcal{G}_k$ satisfies the junction tree condition.

We will now prove the last part of the theorem. Since $\mathbf{x}(\mathcal{G}_1), \ldots, \mathbf{x}(\mathcal{G}_g)$ is a partition of $\{\mathbf{x}_1, \ldots, \mathbf{x}_N\}$, none of the local domains of $\mathcal{G}$ involve any cross terms between $\mathbf{x}(\mathcal{G}_\ell)$ and $\mathbf{x}(\mathcal{G}_k)$ for any $\ell \neq k$. Therefore the global kernel $\beta$ can be written as

$$\beta(\mathbf{x}_1, \ldots, \mathbf{x}_N) = f_1(\mathbf{x}(\mathcal{G}_1)) + f_2(\mathbf{x}(\mathcal{G}_2)) + \cdots + f_g(\mathbf{x}(\mathcal{G}_g)),$$

where, for $\ell = 1, \ldots, g$, $f_\ell(\mathbf{x}(\mathcal{G}_\ell))$ is the sum of the local kernels of all the vertices of $\mathcal{G}_\ell$. Let $v$ be any vertex of $\mathcal{G}$ and let it belong to the $k^{th}$ subtree of $\mathcal{G}$. Let $\sigma_v$ be the state of the vertex $v$ after running the GDL all-vertex message-passing algorithm on $\mathcal{G}$, and $\sigma'_v$ be the state of the vertex after running the GDL all-vertex message-passing algorithm on $\mathcal{G}_k$ only. From the discussion in Section II-B, $\sigma_v$ is the $\mathbf{x}_{\mathcal{I}_v}$-marginalization of $\beta$, and $\sigma'_v$ is the $\mathbf{x}_{\mathcal{I}_v}$-marginalization of $f_k$. We have

$$\sigma_v(\mathbf{x}_{\mathcal{I}_v}) = \min_{\mathbf{x}_{\mathcal{I}_v^c}} \beta = \min_{\mathbf{x}_{\mathcal{I}_v^c}} \sum_{\ell=1}^{g} f_\ell(\mathbf{x}(\mathcal{G}_\ell)).$$

Since each of $f_1, \ldots, f_g$ is a function of disjoint sets of variables, the min and the summation in the above equation can be interchanged. Observing that for all $\ell \neq k$, $\mathbf{x}_{\mathcal{I}_v^c} \cap \mathbf{x}(\mathcal{G}_\ell) = \mathbf{x}(\mathcal{G}_\ell)$, we have

$$\sigma_v(\mathbf{x}_{\mathcal{I}_v}) = \sum_{\ell=1}^{g} \min_{\mathbf{x}_{\mathcal{I}_v^c} \cap \mathbf{x}(\mathcal{G}_\ell)} f_\ell(\mathbf{x}(\mathcal{G}_\ell))$$
$$= \min_{\mathbf{x}_{\mathcal{I}_v^c} \cap \mathbf{x}(\mathcal{G}_k)} f_k(\mathbf{x}(\mathcal{G}_k)) + \sum_{\ell \neq k} \min_{\mathbf{x}(\mathcal{G}_\ell)} f_\ell(\mathbf{x}(\mathcal{G}_\ell))$$
$$= \sigma'_v(\mathbf{x}_{\mathcal{I}_v}) + \sum_{\ell \neq k} a_\ell,$$

where $a_\ell$ denotes the real number $\min_{\mathbf{x}(\mathcal{G}_\ell)} f_\ell(\mathbf{x}(\mathcal{G}_\ell))$. Thus, for any vertex $v$ of $\mathcal{G}$, the functions $\sigma_v$ and $\sigma'_v$ differ only by a scalar. Therefore the solution to $\mathbf{x}_{\mathcal{I}_v}$ obtained from $\sigma'_v$ is

$$\arg\min \sigma'_v(\mathbf{x}_{\mathcal{I}_v}) = \arg\min \left( \sigma_v(\mathbf{x}_{\mathcal{I}_v}) - \sum_{\ell \neq k} a_\ell \right)$$
$$= \arg\min \sigma_v(\mathbf{x}_{\mathcal{I}_v}),$$

which is the solution obtained from $\sigma_v$, and hence is the ML solution. This completes the proof. ∎

## APPENDIX B
## PROOF OF THEOREM 2

In order to prove this theorem we categorize all STBCs into three classes: *(i)* multigroup decodable, *(ii)* conditionally multigroup decodable, and *(iii)* codes in which all the symbols are mutually interfering, which we will call *fully-interfering* STBCs. For $g$-group decodable codes the CML decoder splits into $g$ independent CML decoders, one for each of the $g$ subcodes. Note that each subcode itself can be either conditionally multigroup decodable or fully-interfering. For multigroup and conditionally multigroup decodable codes $\mathcal{O}_{\mathsf{CML}} < |\mathcal{C}|$. For fully-interfering codes CML reduces to brute-force decoding and hence $\mathcal{O}_{\mathsf{CML}} = |\mathcal{C}|$.

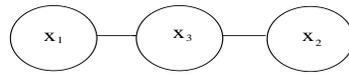

Fig. 22. Moral graph of the smallest conditionally multigroup decodable code.

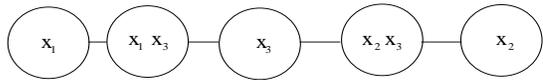

Fig. 23. A junction tree for the smallest conditionally multigroup decodable code.

For each of the three classes of codes we now show that $\mathcal{O}_{\mathsf{GDL}} \leq \mathcal{O}_{\mathsf{CML}}$. For a fully-interfering STBC the junction tree in Section III-B can be used. The complexity of this GDL decoder is of the order of $|\mathcal{C}| = \mathcal{O}_{\mathsf{CML}}(\mathcal{C})$. Since this decoder is only one instance of (possibly) several GDL algorithms for ML decoding this code, we have $\mathcal{O}_{\mathsf{GDL}}(\mathcal{C}) \leq \mathcal{O}_{\mathsf{CML}}(\mathcal{C})$.

Now consider a $g$-group decodable code. The complexity of a CML decoder is sum of the CML complexities of the $g$ subcodes. As explained in Section IV-B, this code can be GDL decoded using a disjoint union of $g$ junction trees, one tree corresponding to each of the $g$ subcodes. Thus, the complexity of GDL decoding is sum of the complexities of GDL decoding each of the $g$ subcodes. Since the subcodes can be either conditionally multigroup decodable or fully-interfering, we only need to show that the theorem is true for conditionally multigroup decodable codes and fully-interfering codes in order to prove the theorem for $g$-group decodable codes. We have already proved the result for fully-interfering codes. In the remaining part of the proof we show that $\mathcal{O}_{\mathsf{GDL}} \leq \mathcal{O}_{\mathsf{CML}}$ for all conditionally multigroup decodable codes.

The proof for conditionally multigroup decodable codes is via induction on $N$, the number of encoding groups of the STBC. The smallest $N$ for which such a code exists is 3 and its corresponding moral graph is shown in Fig. 22. The conditional ML decoder for this code operates with $\Gamma = \{1, 2\}$ and its complexity order is $|\mathcal{A}_3| \max\{|\mathcal{A}_1|, |\mathcal{A}_2|\}$. To decode this code using GDL we can use the junction tree given in Fig. 23. The complexity order of this junction tree equals $|\mathcal{A}_3| \max\{|\mathcal{A}_1|, |\mathcal{A}_2|\} = \mathcal{O}_{\mathsf{CML}}(\mathcal{C})$. Thus we have shown that $\mathcal{O}_{\mathsf{GDL}} \leq \mathcal{O}_{\mathsf{CML}}$ for $N = 3$.

We now prove the induction step. Assume that the theorem is true for all conditionally multigroup decodable codes for which the number of encoding groups is less than $N$. We will now show that the result is true when the number of encoding groups is $N$ as well. Consider a CML decoder with complexity order $\mathcal{O}_{\mathsf{CML}}(\mathcal{C})$ for a code $\mathcal{C}$ with $N$ variables. Suppose this decoder uses $\Gamma \subsetneq \{1, \ldots, N\}$. Let the subcode generated by $\mathbf{x}_\Gamma$ be $g$-group decodable, i.e., let $\mathcal{C}$ be conditionally $g$-group decodable for this choice of $\Gamma$. If the $g$ conditional groups are $\Gamma_1, \ldots, \Gamma_g$, then the complexity order of this CML decoder is $\mathcal{O}_{\mathsf{CML}}(\mathcal{C}) = |\mathcal{A}_{\Gamma^c}| \max_{k=1}^{g} \mathcal{O}_{\mathsf{CML}}(\mathcal{C}_k)$, where $\mathcal{C}_k$ is the subcode generated by the variables $\{\mathbf{x}_n | n \in \Gamma_k\}$. To complete the proof of this theorem it is enough to construct a junction tree for this code whose complexity order is at the most $\mathcal{O}_{\mathsf{CML}}(\mathcal{C})$. For $k = 1, \ldots, g$, the code $\mathcal{C}_k$ is either fully-interfering or conditionally multigroup decodable, and the number of encoding groups in $\mathcal{C}_k$ is less than $N$. Then there exists a GDL decoder for $\mathcal{C}_k$ whose complexity order is upper bounded by $\mathcal{O}_{\mathsf{CML}}(\mathcal{C}_k)$. Let $T_k$ denote the junction tree core for this GDL decoder. Construct a tree $T'_k$ from $T_k$ by appending the variable list $\mathbf{x}_{\Gamma^c}$ to the local domain of every vertex of $T_k$. We now construct a core $T$ using $T'_1, \ldots, T'_g$ and one additional vertex with local domain $\mathbf{x}_{\Gamma^c}$. For every $k = 1, \ldots, g$, arbitrarily choose a vertex of $T'_k$ and connect it to the $\mathbf{x}_{\Gamma^c}$ vertex using a single edge.

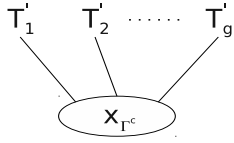

Fig. 24. The tree $T$ in the proof of Theorem 2.

It is straight forward to prove that $T$ is a valid junction tree core for ML decoding of the STBC $\mathcal{C}$. For every vertex in $T'_k$ the local domain size is upper bounded by $|\mathcal{A}_{\Gamma^c}|\mathcal{O}_{\mathsf{GDL}}(\mathcal{C}_k)$. Therefore,

$$\mathcal{O}_{\mathsf{GDL}}(\mathcal{C}) \leq \max_{k=1}^{g} |\mathcal{A}_{\Gamma^c}| \; \mathcal{O}_{\mathsf{GDL}}(\mathcal{C}_k)$$
$$\leq \max_{k=1}^{g} |\mathcal{A}_{\Gamma^c}| \; \mathcal{O}_{\mathsf{CML}}(\mathcal{C}_k) = \mathcal{O}_{\mathsf{CML}}(\mathcal{C}).$$

This completes the proof. ∎

## APPENDIX C
## THE CML AND GDL DECODING COMPLEXITIES OF FULLY-INTERFERING STBCS

The CML algorithm for a fully-interfering code reduces to a brute-force search

$$(\hat{s}_1, \ldots, \hat{s}_{Nt}) = \arg\min f(s_1, \ldots, s_{Nt})$$
$$= \arg\min \sum_{i=1}^{Nt} \left( s_i \xi_i + s_i^2 \xi_{i,i} \right) + \sum_{i<j} s_i s_j \xi_{i,j}.$$

For each of the $q^{Nt}$ values that $(s_1, \ldots, s_{Nt})$ jointly assume, there are $Nt$ terms of type $s_i \xi_i + s_i^2 \xi_{i,i}$ to be computed, and each such term involves $4$ operations. There are $\binom{Nt}{2}$ terms of the type $s_i s_j \xi_{i,j}$ and each term involves $2$ operations. Taking into account the process of summing up these individual terms, the total number of operations in computing $f$ for a given $(s_1, \ldots, s_{Nt})$ is $3\binom{Nt}{2} + 5Nt - 1$. Finding $\arg\min$ of the resulting $q^{Nt}$ values of $f$ takes further $(q^{Nt} - 1)$ operations. Thus, the CML decoding complexity is

$$\mathsf{C}_{\mathsf{CML}}(\mathcal{C}) = q^{Nt}\left(3\binom{Nt}{2} + 5Nt\right) - 1.$$

The GDL decoding of $\mathcal{C}$ involves three steps: computing the kernels $\alpha_n$, $\alpha_{n,m}$, running the GDL message-passing algorithm, and finally the traceback. We use the junction tree of Fig. 1 to decode this STBC. There are $N$ kernels of the type $\alpha_n(\mathbf{x}_n)$. Using the distributive law, $\alpha_n$ can be expressed in terms of $\{s_i\}$ as

$$\alpha_n(\mathbf{x}_n) = \sum_{i \in \psi(n)} s_i \left( \xi_i + s_i \xi_{i,i} \right) + \sum_{i \in \psi(n)} s_i \left( \sum_{\substack{j \in \psi(n) \\ j > i}} s_j \xi_{i,j} \right),$$

where $\psi(n)$ is the set of indices of $\{s_i\}$ that belong to the $n^{th}$ encoding group. The computation of $\alpha_n$ using the above expression involves $q^t(t^2 + 3t)$ operations. There are $\binom{N}{2}$ kernels of the type $\alpha_{n,m}$. Again, with the help of the distributive law, we rewrite $\alpha_{n,m}$ as

$$\alpha_{n,m}(\mathbf{x}_n, \mathbf{x}_m) = \sum_{i \in \psi(n)} s_i \left( \sum_{j \in \psi(m)} s_j \xi_{i,j} \right). \quad (14)$$

The $tq^t$ values of the term $\sum_{j \in \psi(m)} s_j \xi_{i,j}$, one for each pair of $(i, \mathbf{x}_m)$ are precomputed, and then these values are used in (14) to compute $\alpha_{n,m}$. This two step method provides complexity reduction compared to the direct computation of $\alpha_{n,m}$, and can be implemented with $q^{2t}(2t - 1) + q^t(2t^2 - t)$ operations. Using (3), we see that implementing the GDL message-passing schedule takes up $q^{Nt}\binom{N}{2} + q^{2t}N$ operations. Note that the highest order term appearing so far is $q^{Nt}$. The root vertex for the single-vertex GDL and traceback must therefore be chosen in such a way that the complexity of this last step does not contribute to the $q^{Nt}$ term. Choosing any vertex of the type $(\mathbf{x}_n, \mathbf{x}_m)$ will satisfy this requirement as it leads to a traceback complexity of $q^{(N-2)t} + q^{2t} - 2$. Summing up the individual terms, we have the expression for $\mathsf{C}_{\mathsf{GDL}}(\mathcal{C})$ given in (11).

## APPENDIX D
## PROOF OF THEOREM 3

The proof of Theorem 3 is similar to the proof of Theorem 2 given in Appendix B. Here too, we consider three cases: *(i)* multigroup decodable codes, *(ii)* conditionally multigroup decodable codes, and *(iii)* fully interfering codes. From the discussion in Appendix B, we see that it is enough to prove the theorem for fully-interfering codes and conditionally multigroup decodable codes. In Appendix C we have derived the GDL and CML complexities of fully-interfering codes, and the comparison of their leading order terms shows that for such codes $\mathsf{C}_{\mathsf{GDL}}(\mathcal{C}) < \mathsf{C}_{\mathsf{CML}}(\mathcal{C})$.

We now prove the result for conditionally multigroup decodable codes by induction on $N$. The smallest such code involves $N = 3$ encoding groups, and its moral graph is shown in Fig. 22. The CML decoder minimizes

$$\beta = \alpha_3(\mathbf{x}_3) + \alpha_{1,3}(\mathbf{x}_1, \mathbf{x}_3) + \alpha_{2,3}(\mathbf{x}_2, \mathbf{x}_3) + \alpha_1(\mathbf{x}_1) + \alpha_2(\mathbf{x}_2),$$

by conditioning on $\mathbf{x}_3$. For each of the $q^t$ values of $\mathbf{a}_3 \in \mathcal{A}_3$, the CML decoder computes the scalar $\alpha_3(\mathbf{a}_3)$ and the functions $\alpha_{1,3}(\mathbf{x}_1, \mathbf{a}_3)$, $\alpha_{2,3}(\mathbf{x}_2, \mathbf{a}_3)$. It then independently minimizes $\alpha_{1,3}(\mathbf{x}_1, \mathbf{a}_3) + \alpha_1(\mathbf{x}_1)$ and $\alpha_{2,3}(\mathbf{x}_2, \mathbf{a}_3) + \alpha_2(\mathbf{x}_2)$, and finds the conditionally optimal values $\hat{\mathbf{x}}_1(\mathbf{a}_3)$ and $\hat{\mathbf{x}}_2(\mathbf{a}_3)$. From the $q^t$ resulting values of $\beta(\hat{\mathbf{x}}_1(\mathbf{x}_3), \hat{\mathbf{x}}_2(\mathbf{x}_3), \mathbf{x}_3)$, the optimal solution is obtained. The complexity of this algorithm can be shown to be

$$q^{2t}\left(3t^2 + 7t\right) + q^t\left(4t^2 + 3\binom{t}{2} + 5t\right) - 1.$$

The GDL decoder can be implemented on the junction tree shown in Fig. 23. The GDL complexity involves the cost of computing the kernels $\alpha_n$, $n = 1, 2, 3$, $\alpha_{1,3}$ and $\alpha_{2,3}$, running the single-vertex GDL schedule with root vertex $(\mathbf{x}_3)$, and the traceback to find the optimal solution. The complexity of this algorithm is

$$q^{2t}\left(4t + 2\right) + q^t\left(7t^2 + 7t + 3\right) - 3.$$

Comparing the leading terms, we see that the GDL is less complex than the CML decoder. Hence the theorem is true for $N = 3$.

Now consider any conditionally multigroup decodable code with $N \geq 4$ encoding groups, and assume that the theorem is true for all codes with number of encoding groups less than $N$. Assume that the variables corresponding to $\Gamma \subsetneq \{1, \ldots, N\}$ are $g$-group decodable conditioned on the variable list $\mathbf{x}_{\Gamma^c}$. If the $g$ conditional groups are $\Gamma_1, \ldots, \Gamma_g$, the ML metric $\beta$ can be expressed as

$$\sum_{n\in\Gamma^c}\alpha_n + \sum_{\substack{n,m\in\Gamma^c\\n<m}}\alpha_{n,m}+$$

$$\sum_{k=1}^{g}\left[\sum_{\substack{n\in\Gamma_k\\m\in\Gamma^c}}\alpha_{n,m} + \sum_{n\in\Gamma_k}\alpha_n + \sum_{\substack{n,m\in\Gamma_k\\n<m}}\alpha_{n,m}\right].$$

The CML decoder proceeds as follows. For each of the $q^{|\Gamma^c|t}$ values $(\mathbf{a}_n|n \in \Gamma^c) \in \mathcal{A}_{\Gamma^c}$ that the variable list $\mathbf{x}_{\Gamma^c}$ jointly assumes, the CML decoder computes the scalar

$$\sum_{n\in\Gamma^c}\alpha_n(\mathbf{a}_n) + \sum_{\substack{n,m\in\Gamma^c\\n<m}}\alpha_{n,m}(\mathbf{a}_n, \mathbf{a}_m),$$

and the functions $\alpha_{n,m}(\mathbf{x}_n, \mathbf{a}_m)$ for each $n \in \Gamma$ and $m \in \Gamma^c$. It then minimizes the metric

$$\sum_{k=1}^{g}\left[\sum_{\substack{n\in\Gamma_k\\m\in\Gamma^c}}\alpha_{n,m}(\mathbf{x}_n, \mathbf{a}_m) + \sum_{n\in\Gamma_k}\alpha_n(\mathbf{x}_n) + \sum_{\substack{n,m\in\Gamma_k\\n<m}}\alpha_{n,m}(\mathbf{x}_n, \mathbf{x}_m)\right]$$

by multigroup decoding. Minimizing each of the terms corresponding to $k = 1, \ldots, g$ in the above equation is equivalent to decoding the code $\mathcal{C}_k$ generated by $\mathbf{x}_{\Gamma_k}$ by its own CML decoder, and hence each of these terms can be minimized with complexity $\mathsf{C}_{\mathsf{CML}}(\mathcal{C}_k)$. Thus, corresponding to each $\mathbf{a}_{\Gamma^c} \in \mathcal{A}_{\Gamma^c}$ we have a list $\hat{\mathbf{x}}_n(\mathbf{a}_{\Gamma^c})$, $n \in \Gamma$ of conditionally-optimal solutions. Finally, from the $q^{|\Gamma^c|t}$ values of $\beta(\hat{\mathbf{x}}_{\Gamma}(\mathbf{x}_{\Gamma^c}), \mathbf{x}_{\Gamma^c})$, the optimal tuple $(\hat{\mathbf{x}}_{\Gamma}(\mathbf{x}_{\Gamma^c}), \mathbf{x}_{\Gamma^c})$ is chosen. The number of operations involved in this algorithm is given in (15) at the top of the next page. Note that the contribution to the leading order term of $\mathsf{C}_{\mathsf{CML}}(\mathcal{C})$ comes from $q^{|\Gamma^c|t}\sum_{k=1}^{g}\mathsf{C}_{\mathsf{CML}}(\mathcal{C}_k)$.

Let $G_1, \ldots, G_g$ be the junction trees for $\mathcal{C}_1, \ldots, \mathcal{C}_g$ with minimal decoding complexities. Since the number of encoding groups in each of the codes $\mathcal{C}_k$ is less than $N$, the result of this theorem is true for these codes, i.e., $\mathsf{C}_{\mathsf{GDL}}(\mathcal{C}_k) < \mathsf{C}_{\mathsf{CML}}(\mathcal{C}_k)$, for $k = 1, \ldots, g$. We now construct a junction tree for $\mathcal{C}$ using $G_1, \ldots, G_g$. For each $k = 1, \ldots, g$, append the variable list $\mathbf{x}_{\Gamma^c}$ to each of the vertices of $G_k$ and set all the local kernels to zero. From this resulting tree $G'_k$ arbitrarily choose a vertex of type $(\mathbf{x}_n, \mathbf{x}_{\Gamma^c})$, $n \in \Gamma_k$ and connect it to an exterior $(\mathbf{x}_{\Gamma^c})$ vertex by a single edge, as shown in Fig. 25. Set the local kernel at $(\mathbf{x}_{\Gamma^c})$ to zero as well. We now use this tree as the core for the STBC $\mathcal{C}$. For each $n, m \in \Gamma_k$, assign the kernel $\alpha_{n,m}$ to the vertex $(\mathbf{x}_n, \mathbf{x}_m, \mathbf{x}_{\Gamma^c})$ of $G'_k$. For every $n \in \Gamma_k$, assign the kernel $\alpha_n$ to the vertex $(\mathbf{x}_n, \mathbf{x}_{\Gamma^c})$ of $G'_k$. For each pair $n \in \Gamma_k$ and $m \in \Gamma^c$, attach a new vertex $(\mathbf{x}_n, \mathbf{x}_m)$ with kernel $\alpha_{n,m}$ to the vertex $(\mathbf{x}_n, \mathbf{x}_{\Gamma^c})$ of $G'_k$ by a single edge. Attach all the vertices of the type $(\mathbf{x}_n, \mathbf{x}_m)$, $n, m \in \Gamma^c$, with kernel $\alpha_{n,m}$, and all the vertices $(\mathbf{x}_n)$, $n \in \Gamma^c$, with kernel $\alpha_n$, to the $(\mathbf{x}_{\Gamma^c})$ vertex using single edges. It is straightforward to show that this resulting tree $\mathcal{G} = (\mathcal{V}, \mathcal{E})$ is a junction tree for $\mathcal{C}$.

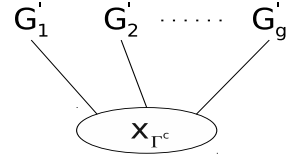

Fig. 25. The tree $T$ in the proof of Theorem 3.

If each of the codes $\mathcal{C}_k$, $k = 1, \ldots, g$, consists of just one encoding group each, then every $G_k$ will consist of just one vertex, and a direct calculation of the number operations involved in GDL decoding using $\mathcal{G}$ shows that $\mathsf{C}_{\mathsf{GDL}}(\mathcal{C}) < \mathsf{C}_{\mathsf{CML}}(\mathcal{C})$. If otherwise, then there exists at least one component $G_k$ with two or more encoding groups. Define $s = \max_{v \in \mathcal{V}} |\mathcal{I}_v|$. Since there is at least one pair of interfering symbols in $\Gamma$, we have $s \geq 2 + |\Gamma^c|$. Let $\mathcal{S}$ be the set of 'largest' vertices in $\mathcal{G}$, i.e., $\mathcal{S} = \{v \in \mathcal{V}|\ |\mathcal{I}_v| = s\}$. Now consider the contribution of each of the three steps: computation of kernels $\alpha_n$ & $\alpha_{n,m}$, running the single-vertex GDL schedule with root $(\mathbf{x}_{\Gamma^c})$, and traceback, to the leading term of $\mathsf{C}_{\mathsf{GDL}}(\mathcal{C})$. The kernels can be computed with the order of complexity $q^{2t}$. The complexity of the GDL single-vertex schedule is of the order of $q^{st}$, and the traceback implementation requires a complexity order less than $q^{st}$. Since $s \geq 2 + |\Gamma^c|$, the only contribution to the leading order term comes from the GDL single-vertex schedule. Recall that $\mathsf{C}_{\mathsf{GDL}}(\mathcal{C}) = \sum_{(u,v)\in\mathcal{E}}(|\mathcal{A}_{\mathcal{I}_u}| + |\mathcal{A}_{\mathcal{I}_v}| - |\mathcal{A}_{\mathcal{I}_u \cap \mathcal{I}_v}|)$. The contribution to the leading order term of $\mathsf{C}_{\mathsf{GDL}}(\mathcal{C})$ comes from the set of all the edges in $\mathcal{E}$ that are incident on the vertices belonging to $\mathcal{S}$. Clearly, every $v \in \mathcal{S}$ belongs to one of the $G'_k$, corresponding to a subcode with two or more encoding groups. From the construction of $\mathcal{G}$, we see that the degree and the edges associated with any vertex from $\mathcal{S}$ in $\mathcal{G}$ are same as the degree and the edges associated with that vertex in the corresponding junction tree $G_k$. It is exactly this set of edges in each $G_k$ that contribute to the leading order terms of $\mathsf{C}_{\mathsf{GDL}}(\mathcal{C}_k)$. Since $\mathcal{G}$ is only one of the many possible junction trees for $\mathcal{C}$, we have $\mathsf{C}_{\mathsf{GDL}}(\mathcal{C}) \leq q^{|\Gamma^c|t}\sum_{k=1}^{g}\mathsf{C}_{\mathsf{GDL}}(\mathcal{C}_k)$, up to the leading order term. From (15) and the assumption made for induction that $\mathsf{C}_{\mathsf{GDL}}(\mathcal{C}_k) < \mathsf{C}_{\mathsf{CML}}(\mathcal{C}_k)$, $k = 1, \ldots, g$, we have

$$\mathsf{C}_{\mathsf{CML}}(\mathcal{C}) = q^{|\Gamma^c|t}\left(\sum_{k=1}^{g}\mathsf{C}_{\mathsf{CML}}(\mathcal{C}_k) + 3\binom{|\Gamma^c|t}{2} + 5|\Gamma^c|t + 2Nt + g\right) - 1. \tag{15}$$

$$\mathsf{C}_{\mathsf{GDL}}(\mathcal{C}) \leq q^{|\Gamma^c|t}\sum_{k=1}^{g}\mathsf{C}_{\mathsf{GDL}}(\mathcal{C}_k)$$
$$< q^{|\Gamma^c|t}\sum_{k=1}^{g}\mathsf{C}_{\mathsf{CML}}(\mathcal{C}_k) \leq \mathsf{C}_{\mathsf{CML}}(\mathcal{C}).$$

This completes the proof. ∎